\documentclass[review,10pt]{elsarticle}
\usepackage{lineno,hyperref}
\modulolinenumbers[5]
\usepackage{amsmath}
\usepackage{amsfonts}
\usepackage{amssymb}
\usepackage{lmodern}
\usepackage{xfrac}
\usepackage{graphicx}
\graphicspath{ {./figures/} }
\usepackage{subfig}
\usepackage{epsfig}
\usepackage{epstopdf}
\usepackage{textcomp}
\usepackage{multirow}
\usepackage{setspace}
\usepackage{fullpage}
\usepackage{float}
\usepackage{caption}
\usepackage{siunitx}
\usepackage{xcolor}
\usepackage{tabularx, booktabs}
\usepackage{enumitem}
\usepackage{dcolumn}% Align table columns on decimal point
\usepackage{bm}% bold math
\usepackage[T1]{fontenc}
\usepackage{gensymb}
\biboptions{sort&compress}

\begin{document}
\begin{frontmatter}

%\preprint{}

\title{\large{\textbf{Orientation selection in alloy dendritic evolution during melt-pool solidification}}}

\cortext[mycorrespondingauthor]{Corresponding author.}
%\cortext[cor1]{\mbox{Corresponding author. \textit{Email address:} \texttt{gsupriyo2004@gmail.com}} (Supriyo Ghosh)}
\author{Saurabh Tiwari}
\author{Supriyo Ghosh\corref{mycorrespondingauthor}}
\ead{supriyo.ghosh@mt.iitr.ac.in; gsupriyo2004@gmail.com}
\address{Metallurgical \& Materials Engineering Department, Indian Institute of Technology Roorkee, Roorkee, UK 247667, India}

%\author[mymainaddress]{\corref{mycorrespondingauthor}}
%\ead{}

%\author{}

%\author[mymainaddress]{\corref{mycorrespondingauthor}}
%\ead{}
%

\date{\today}

\begin{abstract}
Investigations of directionally solidifying melt pools during metal additive manufacturing (AM) reveal that the resulting subgrain cellular structures often grow along crystalline orientations different from the temperature gradient direction, some of which are not even along preferred crystallographic directions. It is well-known that dendrite orientation results from the growth competition between the heat flow direction and preferred crystallographic orientation. Specifically, the competition between interfacial anisotropy and process anisotropy (thermal gradient and interface velocity) during directional solidification leads to rich morphological diversity of the resulting dendritic structures, including tilted dendrites and seaweed patterns. The orientation selection mechanisms of such patterns remain unexplored at high velocity in the frame of AM. This study examines the tilted growth of cellular-dendritic arrays as a function of the misorientation angle ($\theta_R$) between the thermal gradient and crystal lattice directions and other relevant control parameters. We use a phase-field model to explore dendritic evolution in a binary alloy during high-velocity solidification in two-dimensions. We find marked effects of thermal gradient, growth velocity, alloy composition, and anisotropy parameters on the possible growth directions and the primary arm spacing, constitutional undercooling, microsegregation, and secondary phases that arise during dendritic solidification. Our work provides a detailed yet concise presentation on the tilted growth and morphological transition for a broad range of thermal conditions in the high-velocity regime and the full range of $\theta_R$ for establishing orientation selection maps. These results reasonably agree with experimental measurements and should have qualitative relevance for controlling the subgrain structure, chemical segregation, and texture randomization in a wide range of commercial dendrite materials relevant to AM. 
\end{abstract}

\begin{keyword}
Directional solidification \sep Orientation selection \sep Tilted dendrite \sep Seaweed pattern  \sep Phase-field
\end{keyword}

\end{frontmatter}

\section{Introduction}\label{sec_intro}
\begin{figure}[ht]
\centering
\includegraphics[width=\textwidth]{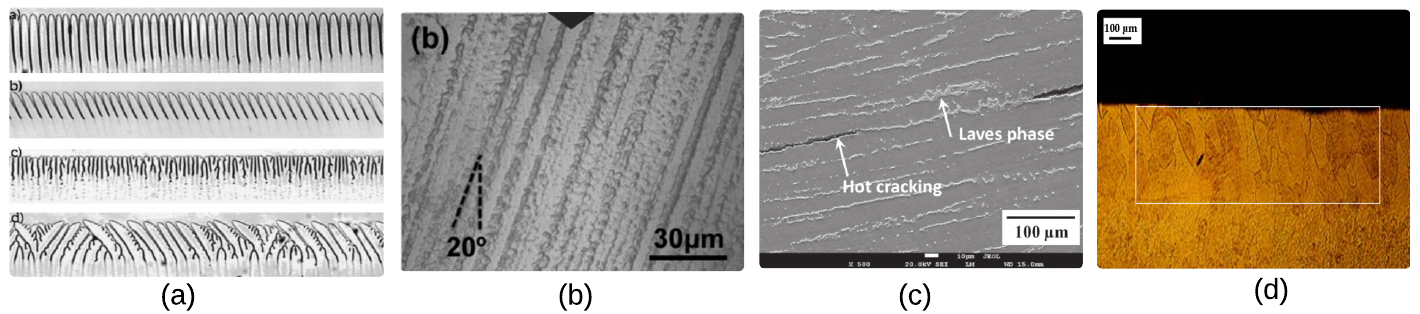}
\caption{Representative experimental observations of dendrite growth directions. (a) From top to bottom, symmetrical dendrites, tilted dendrites, seaweed dendrites, and degenerate dendrites form at low velocity in a transparent metal alloy (reproduced from~\cite{akamatsu1998anisotropy}, with permission from APS). (b) Tilted dendrites with misorientation angle ($\theta_R$) of $\approx$20\degree develop in Inconel 718 in AM (reproduced from~\cite{li2017misorientation}, with permission from Elsevier). (c) The morphology of the secondary phase (\textit{e.g.}, Laves) affects the hot cracking behavior of Inconel 718 (reproduced from~\cite{debroy_additive}, with permission from Elsevier). (d) Possible seaweed growth front develops during laser AM of Ti-6Al 4V (reproduced from~\cite{collins2016microstructural}, with permission from Annual Reviews).}\label{fig_experiment1}
\end{figure}

\begin{figure}[ht]
\subfloat[]{\includegraphics[width=0.52\textwidth]{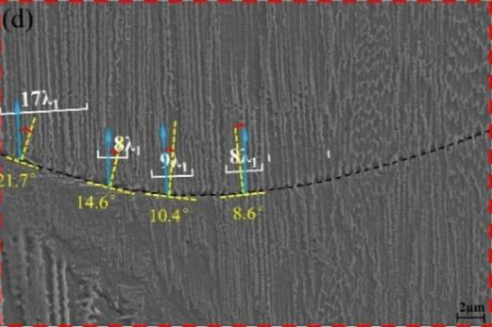}}\hfill
\subfloat[]{\includegraphics[width=0.46\textwidth]{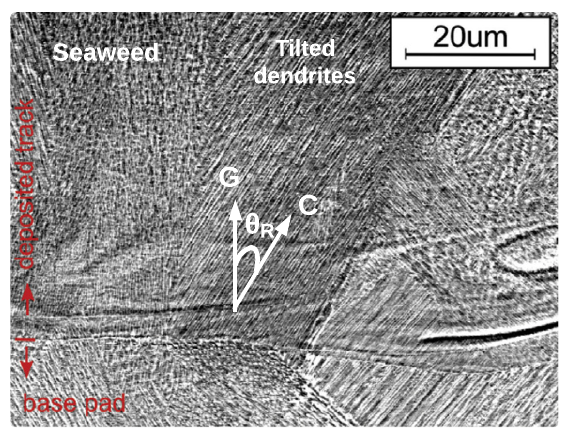}}
\caption{Orientation selection of dendrites in the as-solidified melt-pool (partial) of Inconel 718. The build direction is vertical. (a) Scanning electron microscopy (SEM) image shows the melt-pool boundary angle with respect to $G$, which controls the local misorientation angle ($\theta_R$) and dendrite spacing (reproduced from~\cite{tao2019_kv}, with permission from Elsevier). The blue arrow denotes the build direction, and yellow dashed lines represent the directions normal to the curved isotherm. (b) The disordered seaweed regions with no specific growth direction and the tilted growth regions with a finite $\theta_R$ can be discerned (reproduced from~\cite{acharya2017prediction}, with permission from Elsevier). The transition between these morphologies can be identified by the interplay of location-specific thermal conditions ($G$ and $V$) and $\theta_R$ between the crystalline axis ($C$) and $G$ direction, which affects the degree and orientation of the effective anisotropy in the material.}\label{fig_experiment2}
\end{figure}

Directional solidification of metallic alloys is a well-established process for investigating the competitive growth and interface morphology arising from a non-equilibrium pattern-forming system with technological relevance. As a case study, we consider the example of a selective laser melting process of additive manufacturing (AM) in which the heat source melts the feedstock material to generate a liquid melt pool, which subsequently undergoes a directional solidification process ~\cite{ghosh2023_review,debroy_additive,liu2022additive,review_meltpool,sanchez2021_review,collins2016microstructural}. The solidification front grows locally into the liquid under a given thermal gradient ($G$) and growth velocity ($V$), often resulting in dendritic crystals (Figs.~\ref{fig_experiment1} and~\ref{fig_experiment2}). In the classical treatment of directional configuration, the thermal gradient and interface velocity are aligned onto a principal crystalline axis of the dendrite crystal. Under typical low-velocity conditions, the free energy \textit{anisotropy} of the solid-liquid interface (\textit{i.e.}, interfacial anisotropy) controls the crystallographic orientations in a dendritic structure~\cite{haxhimali2006,akamatsu1998anisotropy}. This anisotropy arises from the cubic lattice structure of the growing solid, with the <100> directions being the preferred growth axes that align with the heat flow direction. However, solidification conditions are often non-uniform along the growth front in the solidifying molten pool (Fig.~\ref{fig_experiment2}a)~\cite{li2017misorientation}. Thus, a competition between the heat flow direction established by the applied thermal gradient and the preferred crystalline direction resulting from interfacial anisotropy sets in (Fig.~\ref{fig_experiment2}b), governing the possible growth directions and rich morphologies of the dendrite crystal~\cite{akamatsu1995symmetry,akamatsu1998anisotropy,li2017misorientation}.

\begin{figure}[ht]
\centering
\includegraphics[scale=0.35]{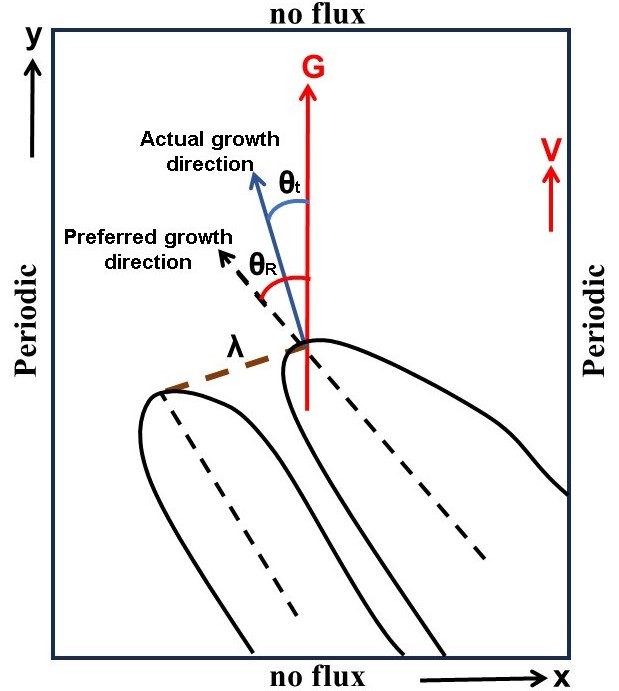}
\caption{Schematic illustration of the misorientation angle ($\theta_R$) and tilt angle ($\theta_t$) during dendritic evolution. The temperature gradient ($G$) and growth velocity ($V$) directions are shown. $\lambda$ is the primary spacing. $G$ is parallel to the growth axis $y$. The boundary conditions (periodic and no-flux) used for simulations are also depicted.}\label{fig_schematic}
\end{figure}

Dendrites grow symmetrically (\textit{i.e.}, axial orientation) during the diffusion-limited growth of isotropic or weakly anisotropic crystals, because constraining the growth to specific orientations other than the temperature gradient is difficult. Symmetrical dendrites can evolve into various orientations, leading to tilted dendrite, seaweed pattern, or even a disordered pattern due to the above growth competition~\cite{akamatsu1995symmetry,akamatsu1998anisotropy,amoorezaei2012,haxhimali2006,utter2001alternating} (Fig.~\ref{fig_experiment1}). To our knowledge, such morphologies and transitions between them are rarely studied at high velocity in the frame of AM. In the literature, the dendrites arising from melt pool solidification are often described by the so-called ``normal-growth rule'', which does not consider the crystal orientation effects~\cite{Trevor2017,karayagiz2020,acharya2017prediction,francois2017}. However, numerical simulations can efficiently examine these effects by changing the solidification front angle between the thermal gradient direction and crystalline axis (Fig.~\ref{fig_experiment2}), which affects the degree and orientation of the crystalline anisotropy and, hence, the growth direction of the dendrites. In view of this, in the following, we will denote by $\theta_R$ the (misorientation) angle between the preferred crystalline orientation and the thermal gradient direction, and we will call the corresponding dendrite orientation the tilt angle $\theta_t$ (Fig.~\ref{fig_schematic}). 

The tilted growth of dendrites at low velocity has been studied extensively through experiments, theory, and modeling~\cite{akamatsu1995symmetry,akamatsu1998anisotropy,amoorezaei2012,haxhimali2006,utter2001alternating,xing2016}. These studies demonstrate that the thermal gradient, growth velocity, and interfacial anisotropy affect the dendrite orientation ($\theta_R$) and morphology. Thus, $\theta_R$ has significant implications for the dendrite characteristics, such as tip profile (\textit{e.g.}, tip speed and radius of curvature), length scale (\textit{e.g.}, primary arm spacing), and compositional variation across the interface (\textit{i.e.}, microsegregation). These observations allow for the identification of the relationships between growth direction and relevant variables, as given by the commonly accepted dendrite growth theories involving $\theta_R$~\cite{dgp_2007,dgp_2008}. Only a few studies~\cite{akamatsu1997similarity,dgp_2008,xing2016,tourret_2015} explore the tilted growth of dendrite arrays in the whole accessible range of 0\degree $\leq$ $\theta_R$ $\leq$ 45\degree for a cubic crystal. It is found that the growth direction changes smoothly with increasing $\theta_R$, with a large enough $\theta_R$ ($>$ 35\degree) often leads to the transition of interface morphology from tilted dendrites to ``seaweed'' patterns, which are characterized by irregular front dynamics with a tip splitting morphology~\cite{akamatsu1995symmetry,utter2001alternating}.
%Dendrites can change directions and morphology with the influence of $\theta_R$ and the above control parameters.

Despite the progress with dendrites, a clear understanding of the effect of varying $\theta_R$ on $\theta_t$ is still lacking, particularly at high velocity. In the frame of AM, $\theta_R$ varies with the location within the melt pool domain~\cite{review_meltpool,tao2019_kv}. Also, a small disturbance in the melt-pool may lead to thermal perturbations at the interface, giving rise to solidification sites with various $\theta_R$ and, hence, various dendrite morphologies and orientations~\cite{li2017misorientation,rannar2017hierarchical}. A coarse columnar structure with large $\theta_R$ typically develops in AM. Such anisotropy-driven tilted dendrites are beneficial for high-temperature creep resistance properties, while the randomly oriented dendrites with their low aspect ratios (\textit{e.g.}, seaweed) can be beneficial for the fatigue and fracture performance of the solidified material. Therefore, understanding the correlation between $\theta_R$ and dendritic evolution is essential in AM. 

The literature on tilted dendrites at high velocity is extremely limited. The majority of these studies do not consider the crystallographic effects, hence, $\theta_R$ = 0\degree is assumed~\cite{Trevor2017,karayagiz2020,acharya2017prediction,francois2017}. However, tilted growth is prevalent in experiments (Fig.~\ref{fig_experiment1}). In addition, growth competition between different morphologies may arise within the melt pool, resulting in tilted and random domains of dendrites (Fig.~\ref{fig_experiment2}). Due to limited research, the orientation selection in such patterns is poorly understood. Numerical simulations of tilted dendrites during electron beam AM of Ti-6Al-4V (adopted as pseudo-binary Ti-10\%X) have been reported in~\cite{chu2020phase}. The effects of $\theta_R$ on the dendritic morphology, compositional variation, and secondary phases in Inconel 718 (approximating as Ni-Nb quasi-binary) have been modeled in~\cite{nie_2014,chen2019am}. However, these studies do not characterize growth dynamics in terms of cell spacing, microsegregation, solute trapping, and constitutional undercooling, which are critical to evaluate the influence of $\theta_R$ on morphological evolution. Also, these studies report limited variations in the thermal and anisotropy parameters, with no interactions among them. We expect such interactions to be more pronounced if the metal is strongly anisotropic and, thus, could play a critical role in the orientation selection at high velocity. 

It is well-known in directional solidification that the interface morphology typically changes in the following overall sequence with increasing velocity (or decreasing thermal gradient): planar $\rightarrow$ cellular $\rightarrow$ dendritic $\rightarrow$ cellular $\rightarrow$ planar. However, the calculation of these growth forms does not consider the anisotropy effects associated with interfacial energy. The orientation selection at a low velocity close to the lower morphological stability limit has been studied extensively~\cite{akamatsu1995symmetry,akamatsu1998anisotropy,amoorezaei2012,haxhimali2006,utter2001alternating,xing2016}. However, limited studies~\cite{chen2019am,chu2020phase,debroy2015texture} investigate the question of orientation selection at high velocity close to the cellular-planar transition instability (\textit{i.e.}, absolute stability). This high-velocity regime is relevant for AM solidification. As described earlier, tilted dendrites are sensitive to morphological instabilities that may lead to the formation of seaweed patterns and other disordered patterns of a different nature from the seaweed pattern, such as degenerate patterns~\cite{akamatsu1995symmetry,akamatsu1998anisotropy,utter2001alternating}. \textit{In situ} characterization of these irregular, random structures is difficult in the final AM microstructure owing to their limited stability and likelihood of forming at the highest $V$ regions (\textit{i.e.}, near the top of the melt-pool); thus, they are likely to get eliminated during post-AM processing. In this study, we attempt to explore the possible dendrite morphologies and orientations that may arise due to different thermal conditions and $\theta_R$ for controlling the grain/subgrain structure and crystallographic texture in AM.

``Numerical experiments'' are an efficient way to address the multivariate problem of orientation selection in dendrites. The phase-field method is primarily used for the computation of dendrites at high velocity~\cite{Trevor2017,karayagiz2020,acharya2017prediction,ghosh_tusas}. This method does not require explicit tracking of the interface; can capture more physical phenomena during the solidification process; and numerical issues associated with the mesh anisotropy and interfacial curvature are handled efficiently. Thus, the phase-field method is arguably the most high-fidelity technique among the microstructure simulation methods used for AM, including the cellular automata, kinetic Monte Carlo, and dendrite needle network~\cite{ghosh2023_review,korner2020modeling,francois2017,tourret2013_dnn}. In this study, we use a phase-field model to simulate the tilted growth of dendrite arrays for various $\theta_R$ and thermal and anisotropy parameters. In the literature, multi-component alloys are often modeled as quasi-binary approximations to simplify physics and implementations and speed up the simulations~\cite{chen2019am,chu2020phase,Trevor2017}. Therefore, we use a quasi-binary adoption of Inconel 718 alloy with Ni-Nb~\cite{knorovsky1989inconel}, with Nb being the most segregating element that controls the formation of the secondary phase (\textit{e.g.}, Laves) and, hence, properties of the solidified material.

The remaining of the article is as follows. In Sec.~\ref{sec_model}, we briefly describe the phase-field solidification model and the material parameters. In Sec.~\ref{sec_results}, we present the simulation results of tilted dendrites that arise due to the variations in different control parameters, including $\theta_R$. We critically analyze the obtained results to demonstrate the influence of relevant variables on dendrite growth directions. We critically discuss the implications of the simulation results and potential future enhancements in Sec.~\ref{sec_discussion}. We summarize and conclude in Sec.~\ref{sec_conclusions}.

%----------------------------
\section{Phase-field model and parameter selection}\label{sec_model}
%\subsection{Phase-Field}\label{sec:phase-field}
Let us consider the directional solidification of a dilute binary alloy that grows dendritically under an externally imposed temperature gradient $G$ along the $y$-axis with a fixed velocity $V$. We have used a quantitative phase-field model to simulate the problem of orientation selection in two-dimensions~\cite{Echebarria2004,karma2001}. The simulations of AM microstructures using this model can be extensively found in the literature~\cite{Trevor2017,acharya2017prediction,chen2019am,Chou_2016,hecht2019am}. Here, we summarize the model equations regarding the implementation of $\theta_R$ in the crystal anisotropy function that triggers the tilted growth of dendrite arrays. 

We must admit that we retain the original interpretation of the relation between the phase-field parameters and the usual sharp interface parameters so that non-equilibrium interface kinetics is present, which is appropriate for solidification at high $V$. Thus, we do not insist on a recovery of local interface equilibrium and use the standard implementation such that a comparative analysis of dendritic evolution for a wide range of $V$ can be made.

As a case study, we simulate a Ni-Nb alloy (as a quasi-binary Inconel 718 alloy) processed \textit{via} selective laser melting. We define the relevant materials and simulation parameters in Table~\ref{tab:parameters}. Our model couples the phase-field ($\phi$) and alloy concentration ($c$) \textit{via} their partial differential equations of motion. The dendritic evolution is given by the time ($t$) evolution of $\phi$ as
\begin{eqnarray}\label{eq_phi}
\tau_0 a(\theta)^2\frac{\partial \phi}{\partial t} = W_{0}^{2} \nabla \cdot \left[{a(\theta)}^2 \nabla\phi\right] - \frac{\partial}{\partial x} \left[ a(\theta) \frac{\partial a(\theta)}{\partial \theta} \frac{\partial \phi}{\partial y} \right] + \frac{\partial}{\partial y} \left[ a(\theta) \frac{\partial a(\theta)}{\partial \theta} \frac{\partial \phi}{\partial x} \right] \\ \nonumber
+ \phi -\phi^3 -  \frac{\lambda}{1-k_e} (1-\phi^2)^2 \left[e^u -1 + \vartheta\right],
\end{eqnarray}
where $\tau_0$ is the phase-field relaxation time, $W_0$ is the interface thickness, and $\theta = \arctan\left(\partial_y\phi/\partial_x\phi\right)$ is the angle between the interface normal ($\hat{n} = \nabla \phi/|\nabla \phi|$) and the $x$-axis, which is normal to the growth direction.

The interfacial energy in cubic crystals is commonly expressed by: $\gamma_{sl} = \bar{\gamma}_{sl} a(\theta)$, where $\bar{\gamma_{sl}}$ is the orientation-average interfacial energy and $a(\theta)$ is the associated anisotropy. When the crystal is rotated with respect to the direction of temperature gradient by angle $\theta_R$, $a(\theta)$ assumes a fourfold anisotropy function given by
\begin{equation}\label{eq_anisotropy}
a(\theta) =  1 + \epsilon_4 \cos \left[ 4 \left(\theta - \theta_R\right)\right],
\end{equation}
where $\epsilon_4$ is the anisotropy coefficient. We choose the reference configuration as $\theta_R$ = 0\degree such that the dendrite orientation of minimal interfacial energy is aligned with temperature gradient or <100> crystallographic direction. 

Since heat diffusion is several orders of magnitude faster than impurity diffusion, the present model neglects latent heat production and a frozen temperature approximation has been imposed to model the temperature field,
\begin{equation}\label{eq_temperature}
\vartheta = (y - y_0 - Vt)\ (1-k_e)/l_T,
\end{equation}
where $y_0$ is the initial interface position. The thermal length $l_T$ is given by $l_T = |m_l| c_0(1-k_e)/(k_e G)$. The dimensionless chemical potential $u$ in Eq.~\eqref{eq_phi} relates to solute concentration $c$ by
\begin{equation}\label{eq:u}
u = \ln \left(\frac{2ck_e/c_0}{1+k_e-(1-k_e)\phi}\right),
\end{equation}
where $k_e$ = $c_s/c_l$ with $c_s$ and $c_l$ are the equilibrium solid and liquid compositions at the interface at a reference temperature $T_0$.

The model solves the solute diffusion equation (without convection) with anti-trapping~\cite{Echebarria2004,karma2001} as
\begin{equation}\label{eq_c}
\frac{\partial c}{\partial t} = -\nabla \cdot \left[ - \frac{1}{2}(1+\phi)\, D_l \, c \, (e^u)^{-1}  \, \nabla e^u +\frac{1}{2\sqrt{2}} W_0 (1-k_e) e^u \frac{\partial \phi}{\partial t} \frac{\nabla\phi}{|\nabla \phi|}\right].
\end{equation}

Equations~\eqref{eq_phi} and~\eqref{eq_c} are rendered dimensionless using the coupling constant $\lambda$, which characterizes the length scale $W_0 = d_0 \lambda /a_1$ and time scale $\tau_0 = (d_0/D)a_2 \lambda^3$ of the model in the thin-interface limit, with $a_1 = 0.8839$ and $a_2 = 0.6267$ ~\cite{Echebarria2004,karma2001}. Our model also includes compositional noise with the magnitude ($F_u^0$ = $10^{-5}$ in~\cite{karma_sidebranching}) at the interface to promote sidebranching, but for brevity, we do not present the details but can be found in~\cite{karma_sidebranching}.

\begin{table}[htbp]
\begin{center}
\setlength{\tabcolsep}{4pt}
    \begin{tabular}{l c r} \hline
      Parameter & Description & Value(s) \\ \hline
      $k_e$&Equilibrium partition coefficient& $0.48$ \\
      $m_l$ & Liquidus slope (K/wt\%) & $-10.5$ \\
      $D$ & Liquid diffusivity (m$^2$/s) & $3 \times 10^{-9}$\\
      $\Gamma$ & Gibbs-Thomson Coefficient (K m) & $3.65 \times 10^{-7}$ \\
      $\Delta T_0$ & Freezing range (K) & 57\\
      $d_0$ & Capillary length (m) &$6.4\times10^{-9}$ \\
      $W_0$ & Interface thickness (m) & $5\times10^{-9}$ \\ 
      $\tau_0$ &  Relaxation  time (s) &  $3.6\times10^{-9}$ \\
      $\lambda$ & Coupling constant &  0.7 \\
      \hline
      $c_0$ & Alloy composition (wt\%) & \{2.5, 5, 7.5\} \\
      $\epsilon_4$& Anisotropy strength & \{0.0003, 0.003, 0.03\} \\
      $G$ & Thermal gradient (K/m) &  \{$10^5$, $10^7$\} \\
      $V$ & Growth rate (m/s) & \{$0.01$, $0.05$, $0.1$, $0.5$\} \\
      $\theta_R$ & Misorientation angles (\degree) & \{0\degree, 15\degree, 30\degree, 45\degree\}
\\ \hline \hline
    \end{tabular}
    \caption{Model alloy and process parameters used in simulations.}
    \label{tab:parameters}
    \end{center}
\end{table}

We solve equations~\eqref{eq_phi} and~\eqref{eq_c} in two-dimensions in a computational domain of $756W_0 \times 2048W_0$ (3024 nm $\times$ 8192 nm) using an explicit Euler finite difference scheme and a standard nine-point stencil operator, with a uniform grid spacing of 0.8$W_0$ and time step of 0.08$\tau_0$. We use a periodic boundary condition for both $\phi$ and $c$ at the left and right boundaries (parallel to the growth axis) and a no-flux condition at the top and bottom boundaries (Fig.~\ref{fig_schematic}). We use a extremely small value of $W_0$ (= 5 nm) to ensure the resulting growth patterns are independent of $W_0$. We performed systematic numerical convergence studies in our previous publications~\cite{ghosh2017primary,ghosh20183d}, hence not shown here. The initial simulation setup assumes a thin solid layer of height 40$W_0$ set at the bottom and a noise amplitude of $\pm$ 0.05. The value of $\phi$ is set to 1 (-1) in the solid (liquid). The initial solid composition is $k_ec_0$, and the liquid composition is $c_0$. Each simulation is initialized with a finite rotation angle $\theta_R$, noting that $\theta_R$ = 45$^\circ$ is the largest misorientation possible for the fourfold symmetry. We analyze the evolution of dendritic morphology and growth direction with respect to the control parameters ($G$, $V$, $\theta_R$, $\epsilon_4$, and $c_0$), as given in Table~\ref{tab:parameters}. Please note that the ranges of solidification conditions ($G$ and $V$) are estimated from macroscopic finite element analysis simulations of melt-pool generated during typical laser melting AM processes, as demonstrated in authors' previous publications~\cite{ghosh2017primary,karayagiz2020}, hence not detailed here. The growth axis ($y$) corresponds to a solidification distance of at least $300D/V$ lengths, allowing for long-time-scale dynamics of dendrite growth upwards into the undercooled melt. We note that lower values of $W_0$ and larger values of $\theta_R$ are computationally very costly.

\section{Results}\label{sec_results}
%Effects of misorientation angle and thermal conditions
\subsection{Effects of misorientation angle and solidification parameters}
\begin{figure}[ht]
\centering
\subfloat[]{\includegraphics[width=0.5\textwidth]{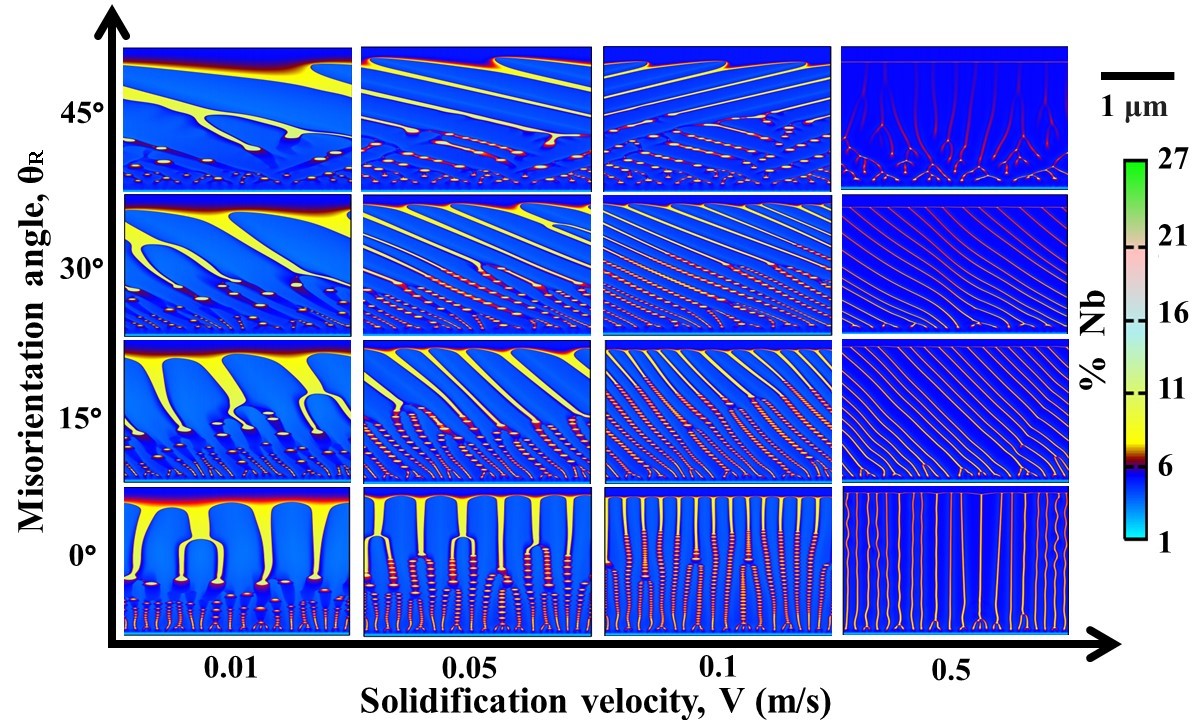}}\hfill
\subfloat[]{\includegraphics[width=0.5\textwidth]{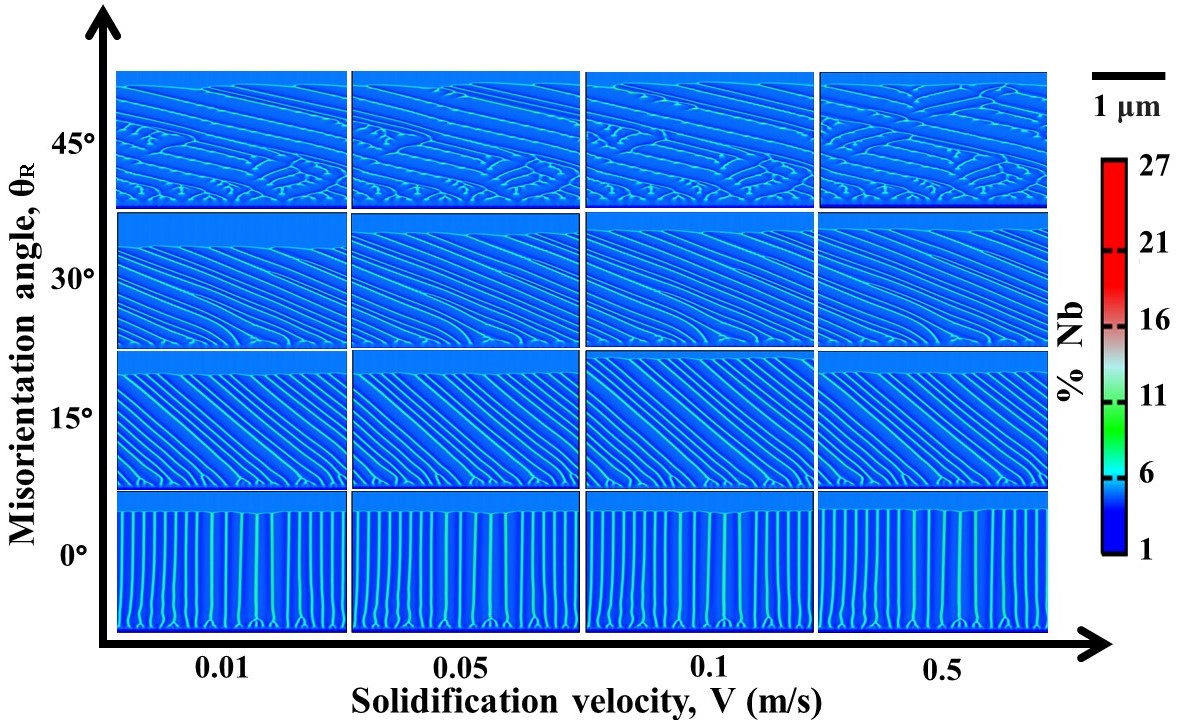}}
\caption{Morphological diagram for dendritic evolution in the space of ($V$, $\theta_R$). The $G$ is fixed at (a) $10^7$ K/m and (b) $10^5$ K/m. The colorbar depicts the composition profile. As $\theta_R$ increases from 0\degree to 45\degree, symmetrical dendrites, tilted dendrites, and seaweed dendrites primarily result. Seaweed dendrites can be classified as columnar seaweed (growth direction: $G$), tilted seaweed (growth direction: $\theta_R$), and degenerate seaweed lacking growth direction. The growth competition between tilted dendrite and seaweed is evident for $\theta_R$ = 45\degree; when ($G = 10^7$ K/m, $V$ = 0.5 m/s) produces a columnar seaweed, while ($G = 10^5$ K/m, $V$ = 0.5 m/s) produces a disordered seaweed. (For interpretation of the references to color in this figure legend, the reader is referred to the web version of this article.)}\label{fig_morphology}
\end{figure}

\begin{figure}[ht]
\centering
\includegraphics[scale=0.6]{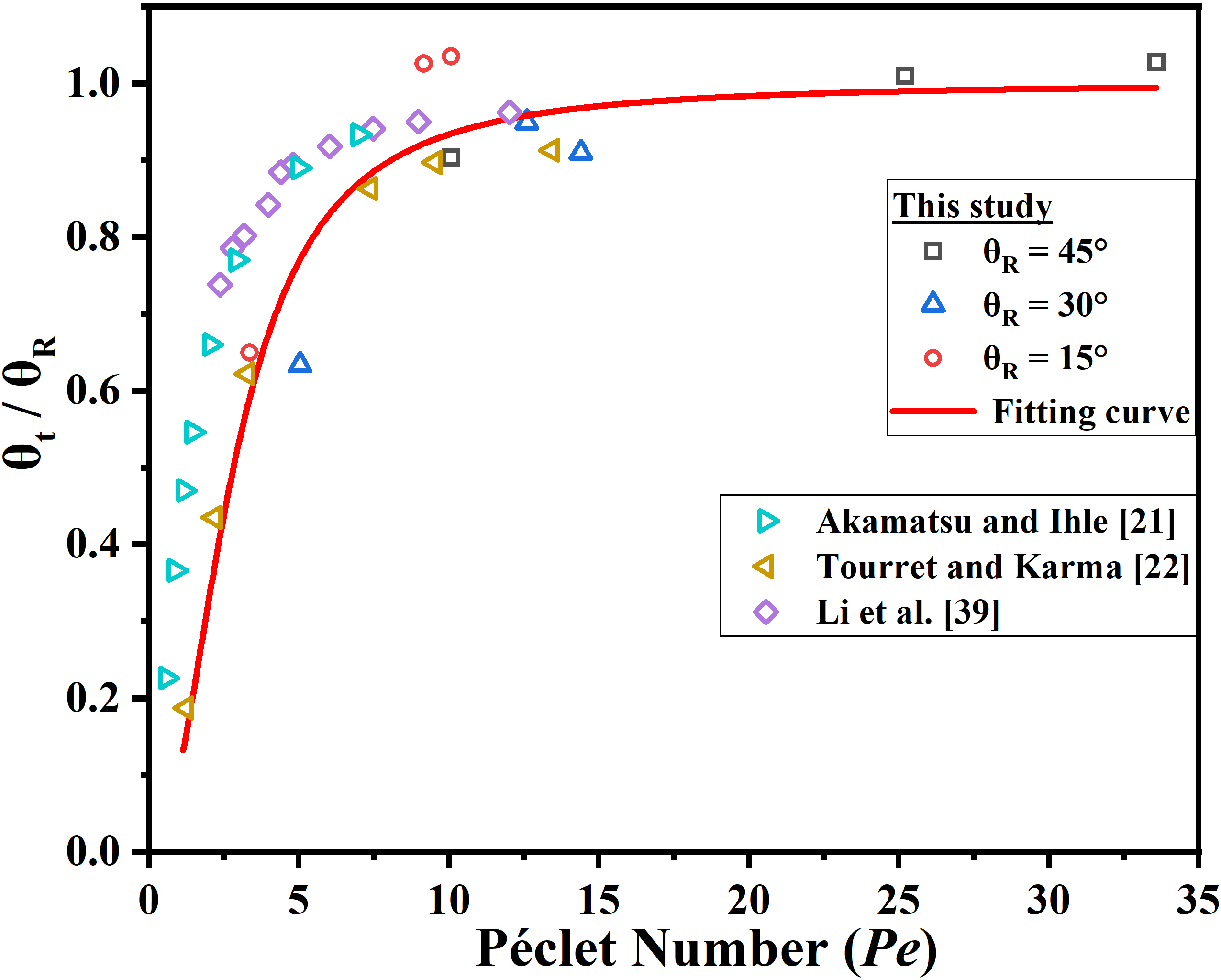}
\caption{The ratio of $\theta_t/\theta_R$ is plotted as a function of P\'eclet number. The plateau signifies the scale invariance. The least-square fit to Eq.~\eqref{eq_dgp} is shown. Here, $G = 10^7$ K/m, $V$ = 0.5 m/s, and $c_0$ = 5\% Nb. We compare our results with the reference literature given in the inset. Note that different materials and processing parameters are used in different studies.}\label{fig_peclet}
\end{figure}

Since we work with a broad parameter space ($c_0$, $\epsilon_4$, $G$, $V$, and $\theta_R$), with their values vary by several orders of magnitude, we present results for selected values for a concise presentation. Unless mentioned, we present data for $G$ = $10^7$ K/m, $\epsilon_4$ = 0.03, and $c_0$ = 5 wt\% (baseline parameters), with varying $V$ $\in$ [0.01 m/s, 0.5 m/s] and $\theta_R$ $\in$ [0\degree, 45\degree] (Table~\ref{tab:parameters}). The results are organized \textit{via} the morphology diagram in Fig.~\ref{fig_morphology}a. First, we describe the morphological evolution for $\theta_R$ = 0\degree. Since $V$ is much larger than the cellular velocity threshold [$V_{cs}$ = $GD/\Delta T_0$ = 0.0005 m/s], the Mullins-Sekerka instability~\cite{Mullins1964} at the initial interface caused by the noise amplifies into cellular/dendritic structures that grow parallel to $G$. As $V$ approaches the absolute velocity [$V_{ab}$ = $\Delta T_0 D/ (k_e \Gamma)$ = 1 m/s], dendrites become progressively finer, with their average tip-to-tip distance ($\lambda$) and the depth of the groove between dendrites continually decrease. With increasing $\theta_R$ for fixed $V$, dendrites often deviate from the $G$ direction to the preferred crystallographic direction, establishing the tilt angle $\theta_t$. In all simulations, $\theta_t$ scales almost linearly with $\theta_R$ (within numerical uncertainty of $\pm$3\degree). It is reported in the literature~\cite{akamatsu1997similarity,dgp_2008} that for $V$ >> $V_{cs}$, $\theta_t$ essentially becomes a function of the spacing P\'eclet number ($Pe = \lambda V/D$). Thus, we plot the $\theta_t$ data with respect to $Pe$ (in Fig.~\ref{fig_peclet}) following the so-called DGP (after Deschamps, Georgelin, and Pocheau) law~\cite{dgp_2007,dgp_2008}:
\begin{equation}\label{eq_dgp}
\frac{\theta_t}{\theta_R} = 1 - \frac{1}{1+\alpha Pe^\beta},
\end{equation}
where $\alpha$ and $\beta$ are the fitting parameters. Previous studies~\cite{tourret_2015} confirm the validity of the DGP law only under similar growth conditions and in model transparent alloys, with typical values of $\alpha$ = 0.5 (0.2) and $\beta$ = 1.6 (1.9) have been reported in experiments~\cite{akamatsu1997similarity} (and simulations~\cite{li2012phase,xing2015phase,liu2022_tilted}). However, the limitations of this theory in the high-anisotropy and high-$V$ regimes have not been studied so far. To test this theory, we first average the $\theta_t$ data obtained from all the simulations of various $\theta_R$ for a given $V$, repeated by varying $V$, and then fit the data to Eq.~\eqref{eq_dgp} to attain $\alpha$ = 0.1 and $\beta$ = 2.1. As expected, reasonable deviations from the reported DGP fitting exist, but the trend of $\theta_t/\theta_R$ is similar to previous studies~\cite{xing2017effect,liu2022_tilted,xing2018growth}. Since changing $V$ will change $\lambda$, we include their combined effects for calculating $Pe$, as typically observed in experiments. However, previous numerical studies~\cite{xing2017effect,liu2022_tilted,xing2018growth} often increase $Pe$ by increasing $V$ for a fixed $\lambda$, which may contribute to different $\alpha$ and $\beta$ compared to ours.

With increasing $\theta_R$ for fixed $V$, the tilted arrays become coarser with the increase of $\lambda$. As $\theta_R$ increases from 0\degree to 45\degree, the dendrites progressively become sensitive to instabilities, leading to the generation of the rightward- and leftward-tilted dendrites (\textit{i.e.}, seaweed) at the transient stages before the transition to tilted dendrites (where the tips are needle-shaped) takes place near the end of the run. The seaweed pattern typically arises from alternating tip splitting instability, in which the tip alternately splits to the left and to the right, resulting in the growth lacking the apparent preferred orientation of the crystal. The morphological transition from seaweed to tilted dendrites suggests that dendrites are dynamically preferred over seaweed for the given growth conditions. Such growth competition is often the precursor to the formation of various irregular dendritic morphologies, including twinned dendrites (``doublons'') and ``degenerate'' seaweeds~\cite{akamatsu1995symmetry,akamatsu1998anisotropy,xing2016}. In the simulation with the highest $V$ (0.5 m/s) and $\theta_R$ (45\degree), a columnar seaweed morphology develops, where the tips are almost flat and the growth closely follows the $G$ direction. Although not shown here, a stable planar interface develops when $V$ exceeds $V_{ab}$. 

The tilted growth pattern, on average, remains approximately the same for a smaller value of $G$ =  $10^5$ K/m under the same baseline growth conditions (Fig.~\ref{fig_morphology}b). For a low $\theta_R$, tilted growth result with $\theta_t$ $\approx$ $\theta_R$. However, a closer inspection of the pattern for $\theta_R$ = 45\degree reveals the existence of transient growth competition between the degenerate seaweed and tilted dendrite, that degree of which clearly increases with increasing $V$. This selection process continues until the tilted dendrites dominate at a low $V$ and the degenerate seaweed at a large $V$. Notably, the interfacial morphology and tip profile are remarkably different under the same condition ($V$ = 0.5 m/s, $\theta_R$ = 45\degree) in Fig.~\ref{fig_morphology}a and~\ref{fig_morphology}b. In Fig.~\ref{fig_morphology}a (\textit{i.e.}, $G = 10^7$ K/m), the interfacial morphology assumes a transitional state between the seaweed and planar fronts, with the growth direction approaching $G$. In contrast, Fig.~\ref{fig_morphology}b (\textit{i.e.}, $G = 10^5$ K/m) develops a degenerate seaweed with no definite growth direction. The orientation selection in these patterns implies that degenerate dynamics could prevent a definite growth direction from emerging during dendrite evolution. Overall, we found that the transition from tilted dendrites to seaweed is favored with increasing $\theta_R$, decreasing $G$, and increasing $V$ ($\to$ $V_{ab}$). 
%To avoid repetition, we do not show morphologies for intermediate $G$, $V$, and $\theta_R$ values.

\subsection{Data analysis}
We examine the effects of $\theta_R$ on the fundamental aspects of dendritic growth, such as the primary dendrite arm spacing ($\lambda$), solute concentration profile across the dendritic-interdendritic regions ($k_V$), radius of curvature of the growth fronts ($\rho_{tip}$), constitutional undercooling of the liquid at the interface ($\Delta T_{tip}$), and the distribution of interdendritic secondary phases in the solidified material ($f_d$). These morphological features could play a critical role in the growth orientation selection of columnar dendrites at high $V$~\cite{ghosh2017primary,Trevor2017,ghosh2018predictive}. Therefore, we characterize the combined effects of varying $(\theta_R$, $V$, $G$, $\epsilon_4$, $c_0$) on the dendrite features that arise during tilted growth. 

\begin{figure}[ht]
\centering
\subfloat[]{\includegraphics[width=0.33\textwidth]{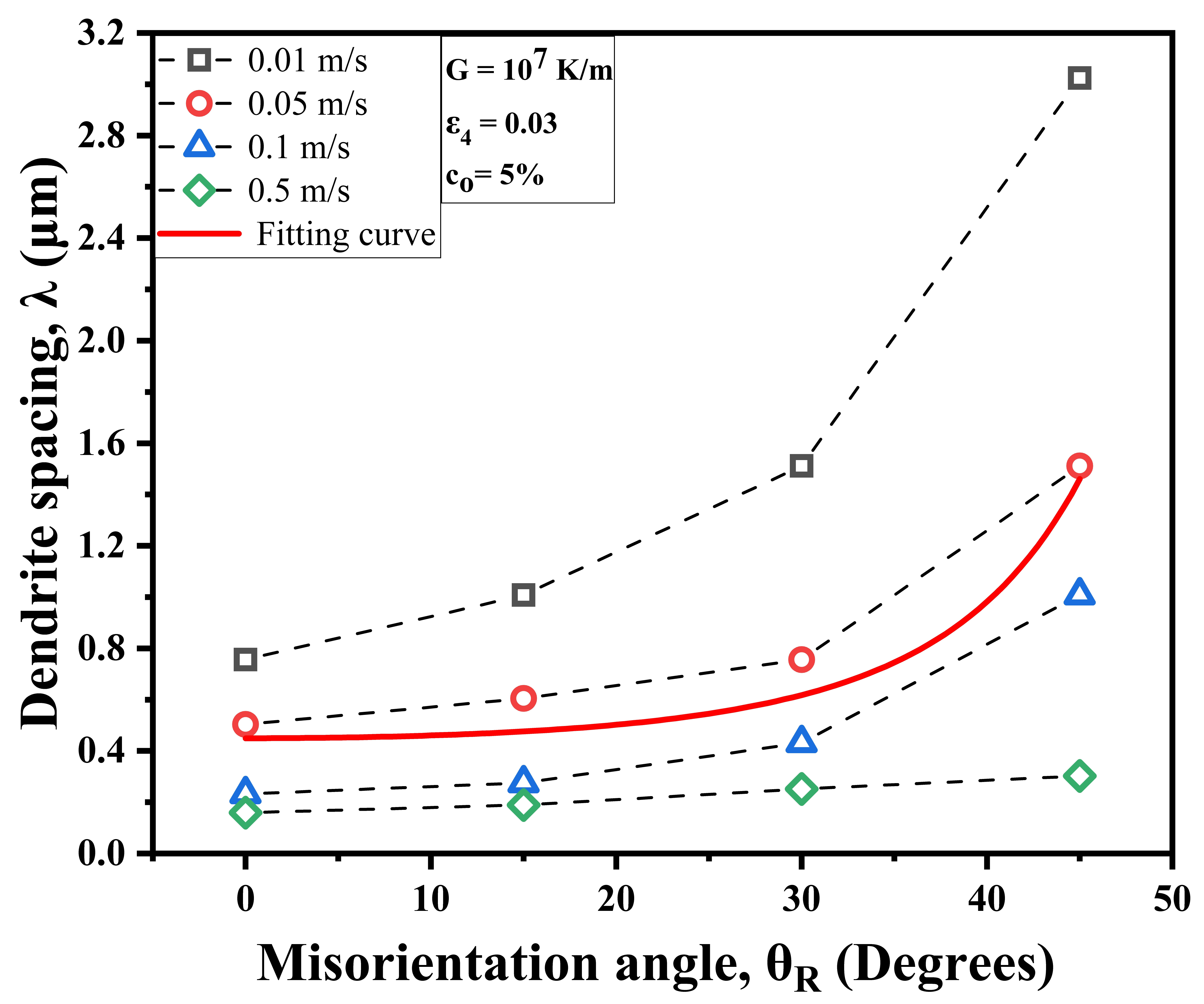}}\hfill
\subfloat[]{\includegraphics[width=0.33\textwidth]{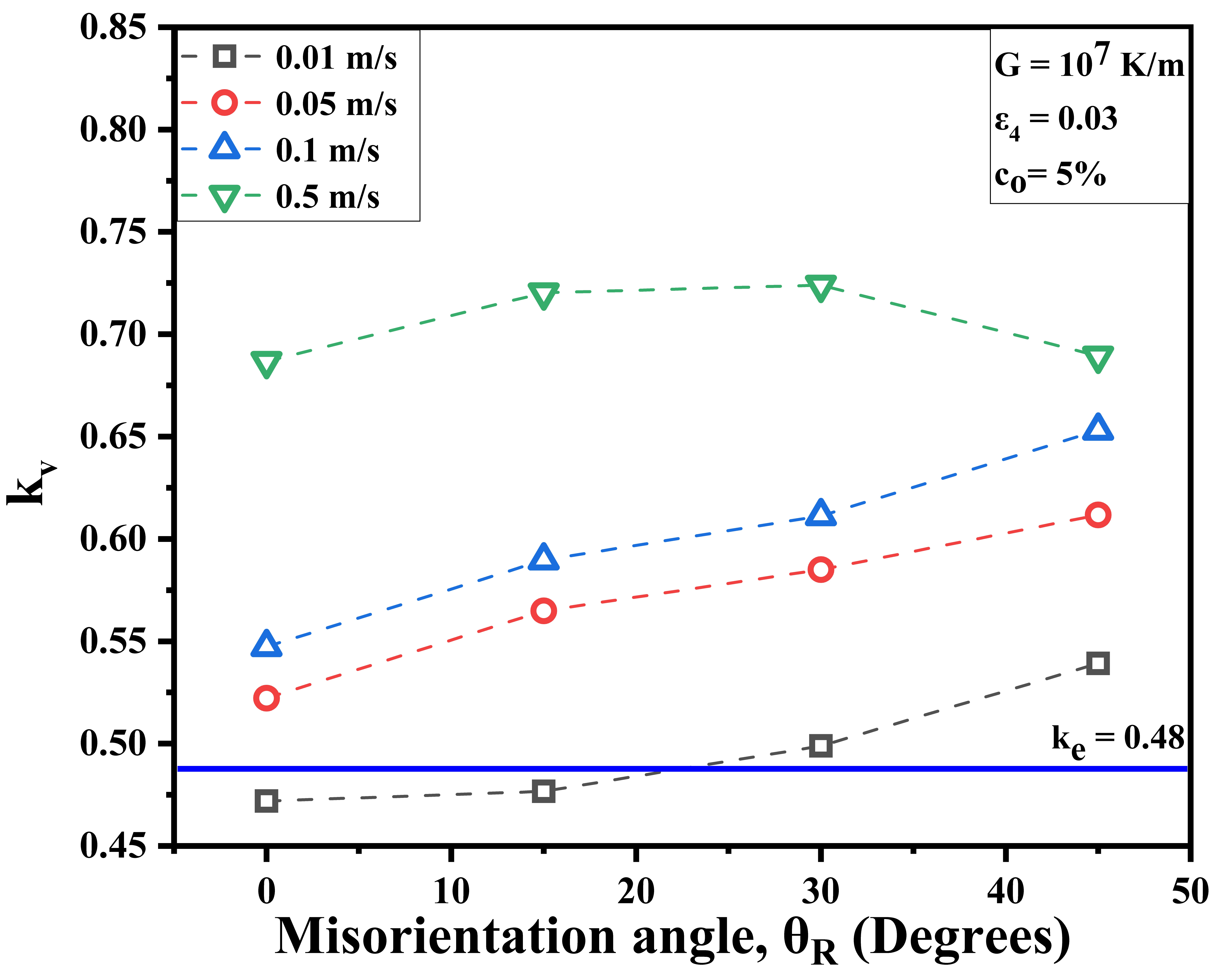}}\hfill
\subfloat[]{\includegraphics[width=0.33\textwidth]{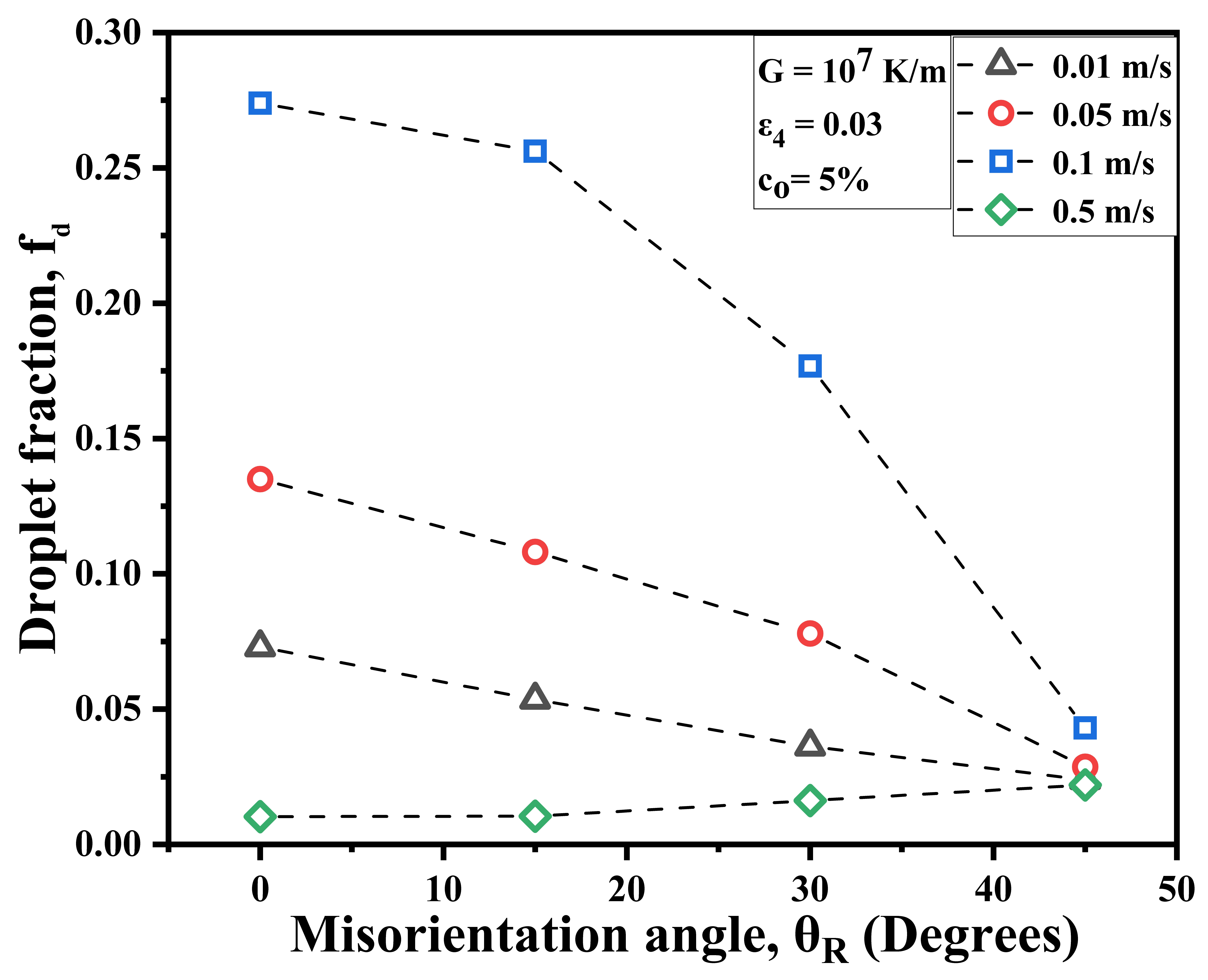}}
\caption{The variations of dendritic features, (a) primary spacing ($\lambda$), (b) segregation ratio ($k_V$), and (c) solute-rich droplet fraction ($f_d$), are shown in the space of ($\theta_R$, $V$) for fixed $G = 10^7$ K/m (correspond to Fig.~\ref{fig_morphology}a). The fixed parameters and variable $V$ are shown in the inset. The fit to Eq.~\eqref{eq_dgp} is shown in Fig.~\ref{fig_features}a.}\label{fig_features}
\end{figure}

Since dendritic growth is irregular (\textit{e.g.}, degenerate seaweed) within some parameter ranges, as a first approximation, we count the number of tips over the entire solidification front and then divide the domain size by this number for a rough estimate of $\lambda$. The results are shown in Fig.~\ref{fig_features}a, where $\lambda$ increases with increasing $\theta_R$, in agreement with previous studies~\cite{gandin1996orientation,tourret_2015,xing2015phase,xing2017effect}. The magnitude of $\lambda$ increases at least by a few times (\textit{i.e.}, the microstructure becomes coarser) over the entire range of $\theta_R$. When the variations in $G$ and $V$ are included, $\lambda$ increases with decreasing $V$ (and increasing $G$) for a given $\theta_R$ as expected. The $\lambda$ data can be compared with the analytical solutions following Hunt~\cite{hunt1979} and Kurz and Fisher~\cite{kurz1980}, albeit for $\theta_R$ = 0\degree, as detailed in~\cite{ghosh2017primary,karayagiz2020}. Gandin et al.~\cite{gandin1996orientation} and later Tourret and Karma~\cite{tourret_2015} incorporated an orientation correction factor $F(\Theta)$ in these relationships that links $\lambda$ to $\theta_R$ in the following power law:
\begin{equation}\label{eq_spacing}
\lambda \propto \Delta T_{0}^{0.25}\, V^{-0.25}G^{-0.5}\, F(\Theta),
\end{equation} 
where $F(\Theta)$ = $1 + d [(\cos \theta_R)^{-e} - 1]$ with the parameters $d$ = 0.15 and $e$ = 8 estimated for organic analog to metals (\textit{i.e.}, succinonitrile-acetone alloys characterized by very small $k_e$ and $D$). Equation~\eqref{eq_spacing} shows that $\lambda$ should increase with increasing $\theta_R$, similar to our observations. The previous studies assume $d$ and $e$ as the fitting parameters, which we estimate as $d$ = 0.2 and $e$ = 7.1. Similar deviations in $d$ and $e$ are commonly observed in previous studies~\cite{xing2018orientation,xing2018growth}. Notably, our results are reasonable when considering the different alloy systems and velocity regimes used in previous studies~\cite{gandin1996orientation, tourret_2015,xing2018orientation,xing2018growth}. In most studies, $\lambda$ is not calculated following the natural evolution of a planar interface, as in our simulations, but rather is set empirically to minimize the relevant parameters in simulations and, hence, the computing time. This could also lead to differences between the literature and our study regarding the interface morphology, particularly when $\lambda$ is scaled with $Pe$ to characterize the tilt angle (Fig.~\ref{fig_peclet}).

\begin{figure}[ht]
\centering
\includegraphics[scale=0.3]{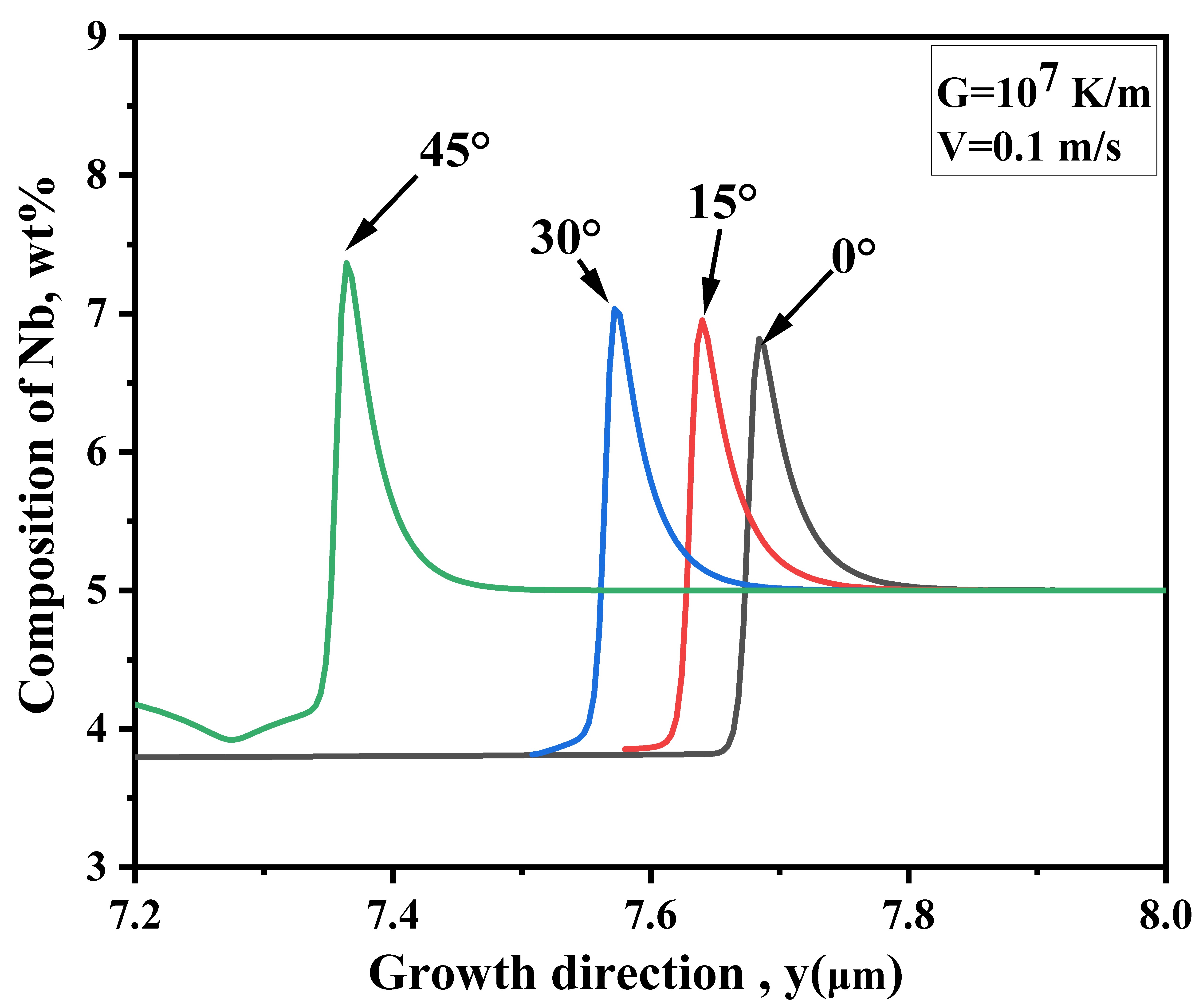}
\caption{Concentration profiles plotted through the leading cell tip for various $\theta_R$ at fixed $G$ and $V$.}\label{fig_solute_profile}
\end{figure}

As the tilted cells grow by rapid freezing, the build-up of solute at the interface due to $k_e<1$ (Table~\ref{tab:parameters}) enriches the interdendritic liquid. We plot the typical solute profiles along a line through the tip for various $\theta_R$ in Fig.~\ref{fig_solute_profile}. We characterize these concentration profiles by the ratio of the solid concentration ($c_s$) just behind the tip to the maximum concentration ($c_{tip}$), taken as a measure of the partition coefficient ($k_V$). The data for 45\degree is noisy due to growth competition and morphological transition between seaweed and tilted dendrites. It is well-known that $k_V$ increases with increasing $V$ (or decreasing $G$) due to the insufficient time available for solute redistribution across the interface~\cite{Boettinger1999,ghosh2017primary}. Thus, we mainly focus here on the effects of varying $\theta_R$ on $k_V$. Since local equilibrium fails at high $V$ due to the details of the diffusion in the liquid, $k_V$ increasingly deviates from $k_e$ with increasing $\theta_R$ (similar to the effects of $V$) (Fig.~\ref{fig_features}b). For reference, we fit the $k_V$ data to the Aziz solute trapping function~\cite{Aziz1982} given by
\begin{equation}\label{eq_aziz}
k_V (V) = \frac{k_e+V/V_D}{1+V/V_D}\, F(\Theta),
\end{equation}
where the fitting parameter $V_D$ is a characteristic velocity for a given alloy related to the magnitude of its solute trapping behavior, thus, independent of $\theta_R$. Therefore, we prepare the data by averaging $k_V$ over all simulations of $\theta_R$ for each $V$, repeated for all $V$, then fit to Eq.~\eqref{eq_aziz} for a $V_D$ = 0.5 m/s. Although not shown here, the estimated $V_D$ differs when the $k_V$ data corresponding to independent $\theta_R$ values are used for the Aziz fitting. This suggests that an orientation correction factor $F(\Theta)$, similar to Eq.~\eqref{eq_spacing}, should be included in Eq.~\eqref{eq_aziz} following a heuristic approach. This correction can be attributed to the observations of increasing $\lambda$ with increasing $\theta_R$. A large $\lambda$ allows for the formation of a wide interdendritic groove to enclose a large amount of solute-rich liquid, increasing $k_V$. Thus, we observe that the maximum solute concentration at the interface in the liquid increases with $\theta_R$ as 6.82 wt\% ($\theta_R$ = 0\degree), 6.95 wt\% ($\theta_R$ = 15\degree), 7.03 wt\% ($\theta_R$ = 30\degree), and 7.37 wt\% ($\theta_R$ = 45\degree), as shown in Fig.~\ref{fig_solute_profile}. Moreover, previous studies~\cite{karayagiz2020,Trevor2017,ghosh2017primary} often consider $\theta_R$ = 0\degree and report that $k_V$ is less sensitive to $G$ at high $V$ close to $V_{ab}$. Here, we find that $k_V$ is quite sensitive to $G$ as well as $V$ for nonzero $\theta_R$ in the high-$V$ regime (Fig.~\ref{fig_effects_G}). 

\begin{figure}[ht]
\centering
\includegraphics[scale=0.4]{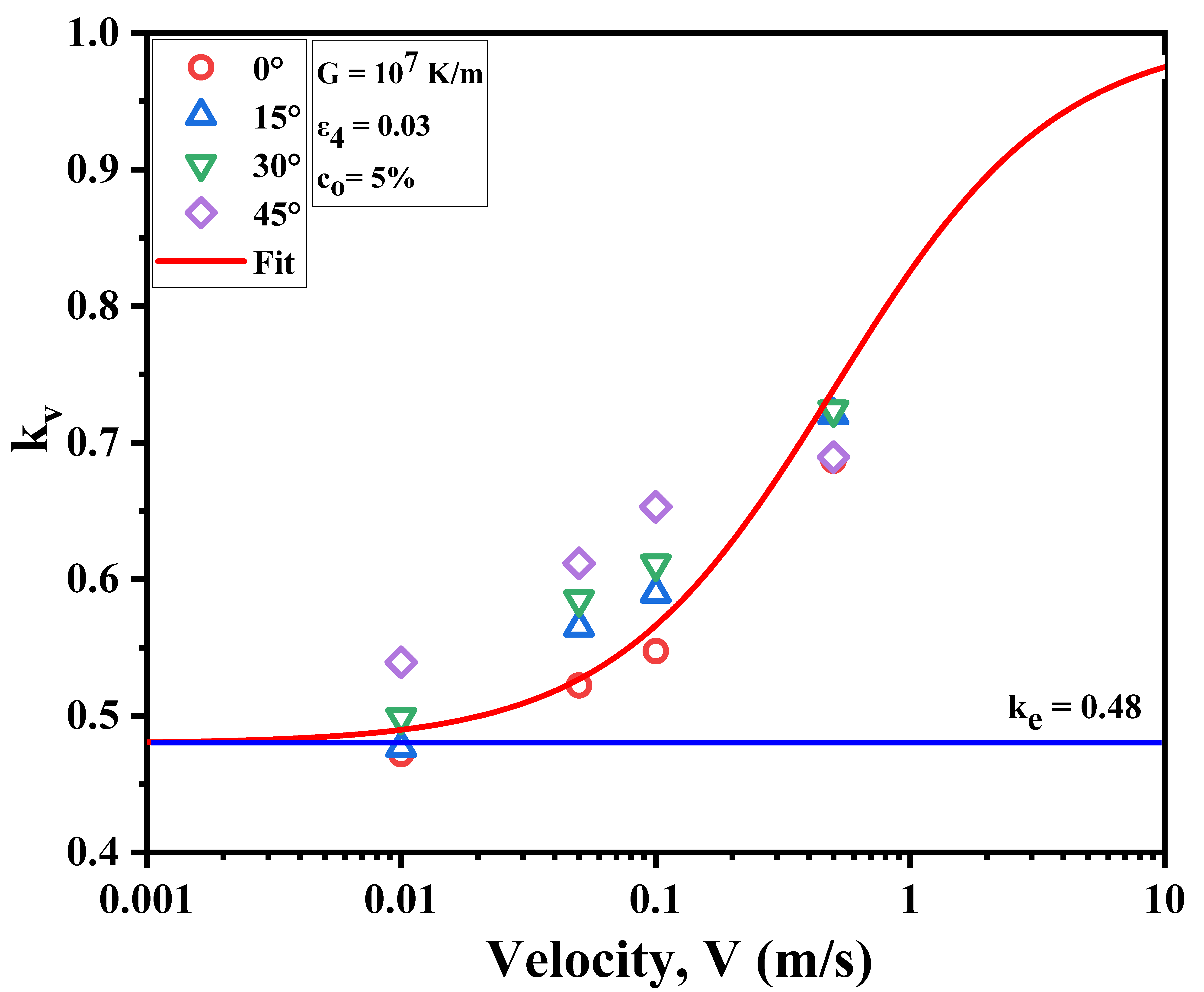}
\caption{Variation of $k_V$ with $V$ for different values of $\theta_R$. We average the $k_V$ data obtained from each $V$ simulation run with four different $\theta_R$, repeated for various $V$, and fit them to Eq.~\eqref{eq_aziz} to obtain $V_D$ = 0.5 m/s.}\label{fig_aziz}
\end{figure}

\begin{figure}[ht]
\centering
\includegraphics[scale=0.4]{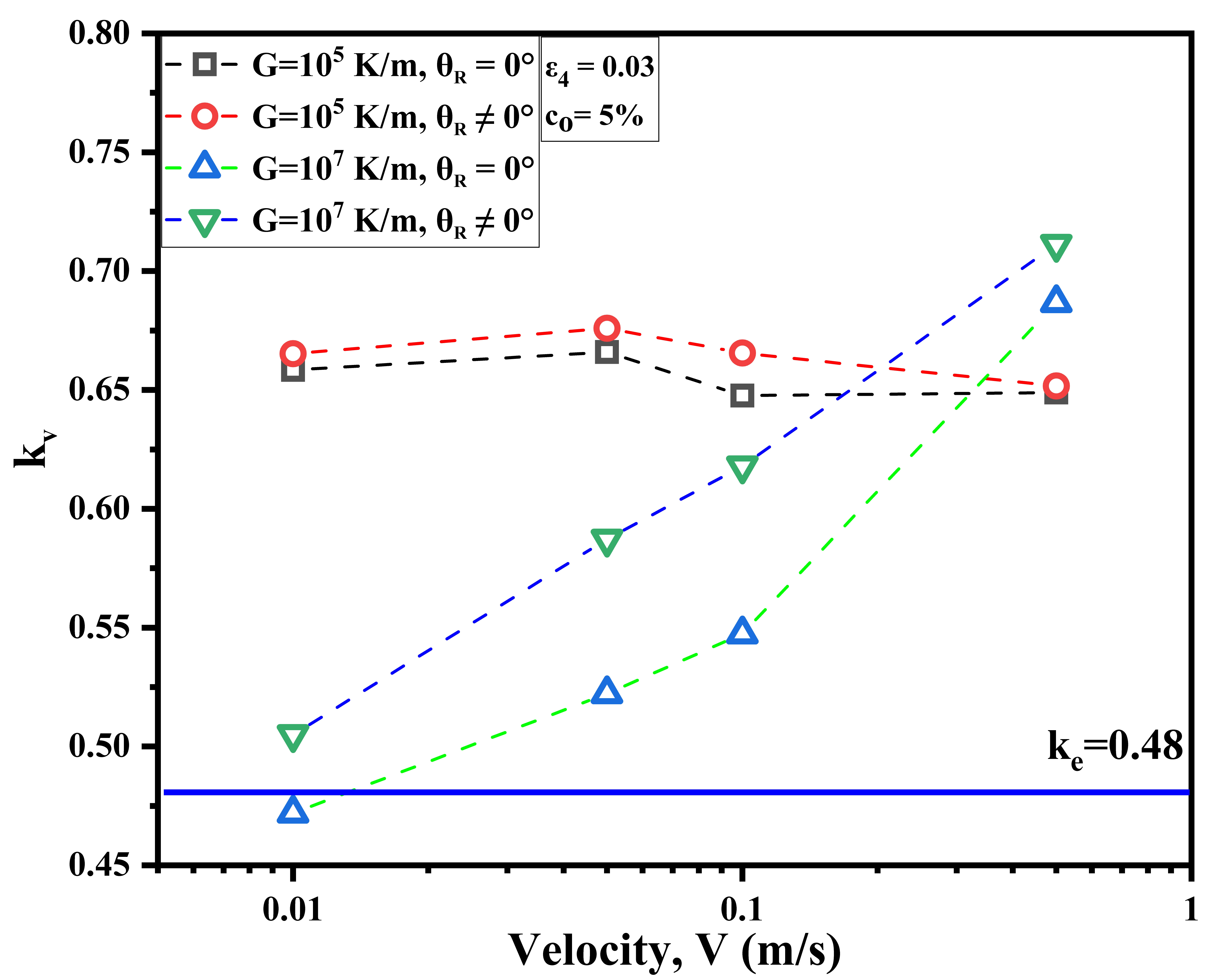}
\caption{The variation in $k_V$ illustrates that the effects of nonzero $\theta_R$ are more pronounced in the high $G$ regime. We plot the average $k_V$ data obtained from simulations with various nonzero $\theta_R$ (\textit{i.e.}, 15\degree, 30\degree, and 45\degree) for a given $G$ and $V$.}\label{fig_effects_G}
\end{figure}

During cellular/dendritic growth, the solute-rich liquid is periodically ``pinched off'' from the root of the interdendritic region, resulting in a trail of liquid (Fig.~\ref{fig_morphology}, $\theta_R$ = 0\degree) or liquid ``droplets'' (Fig.~\ref{fig_morphology}, $\theta_R$ = 45\degree) between the dendrites. Such phenomenon is analogous to the Plateau-Rayleigh instability, as observed previously for axial dendrites with $\theta_R$ = 0\degree~\cite{ungar1985cellular,Ghosh2018_droplets}. Here, we show that such a ``dripping'' phenomenon exists even during the tilted growth and, thus, could be influenced by the variation of $\theta_R$. The melt droplets subsequently freeze, leading to solidification sites with the highest $k_V$, and hence may have consequences for forming non-equilibrium secondary phases (\textit{e.g.}, Laves phase) at the end of solidification. Therefore, the footprint of the droplet network within the dendrite framework could be crucial for the morphological evolution of the secondary phases. The trajectory of the droplet is continuous for symmetrical dendrites at $\theta_R$ = 0\degree and for tilted dendrites at a low $\theta_R$, which could lead to secondary phase formation with a long chain morphology. With increasing $\theta_R$, the trail of the droplets increasingly becomes non-uniform, leading to discrete distributions (or short-chain morphology) of the solid phases. For a large $\theta_R$, tilted dendrite-to-seaweed transition (or vice versa) takes place (Fig.~\ref{fig_morphology}, 45\degree). Such transition arises due to the merging of neighboring dendrites with arbitrary growth directions or when the sidebranches join neighboring dendrites, separating interdendritic liquid into several discrete pockets. As a result, the secondary phases could develop as isolated, discrete particles (or shorter chains). By preventing the formation of long chain morphology, the hot cracking resistance of the solidified material can be significantly improved. Therefore, solidification sites with large $\theta_R$ are often considered beneficial for the mechanical behavior of the solidified material~\cite{nie_2014,chen2019am,li2018effect}. We estimate the secondary phase amount by calculating the area fraction of the droplet with a melt composition exceeding $c_0/k_e$ (\textit{i.e.}, non-equilibrium melt enrichment within the solid). The results in Fig.~\ref{fig_features}c clearly show that the $\theta_R$ affects the expected secondary phase amount ($f_d$) in the solid material. The simulated $f_d$ data agrees with the previous observations of the secondary phase amount that decreased with increasing $\theta_R$ (or $V$)~\cite{tao2019_kv,chen2019am}. In addition, $G$ and $V$ also control the growth mode of the dendrites (Fig.~\ref{fig_morphology}) and, thus, may affect the morphology of the secondary phase particles.

\begin{figure}[ht]
\centering
\subfloat[]{\includegraphics[width=0.33\textwidth]{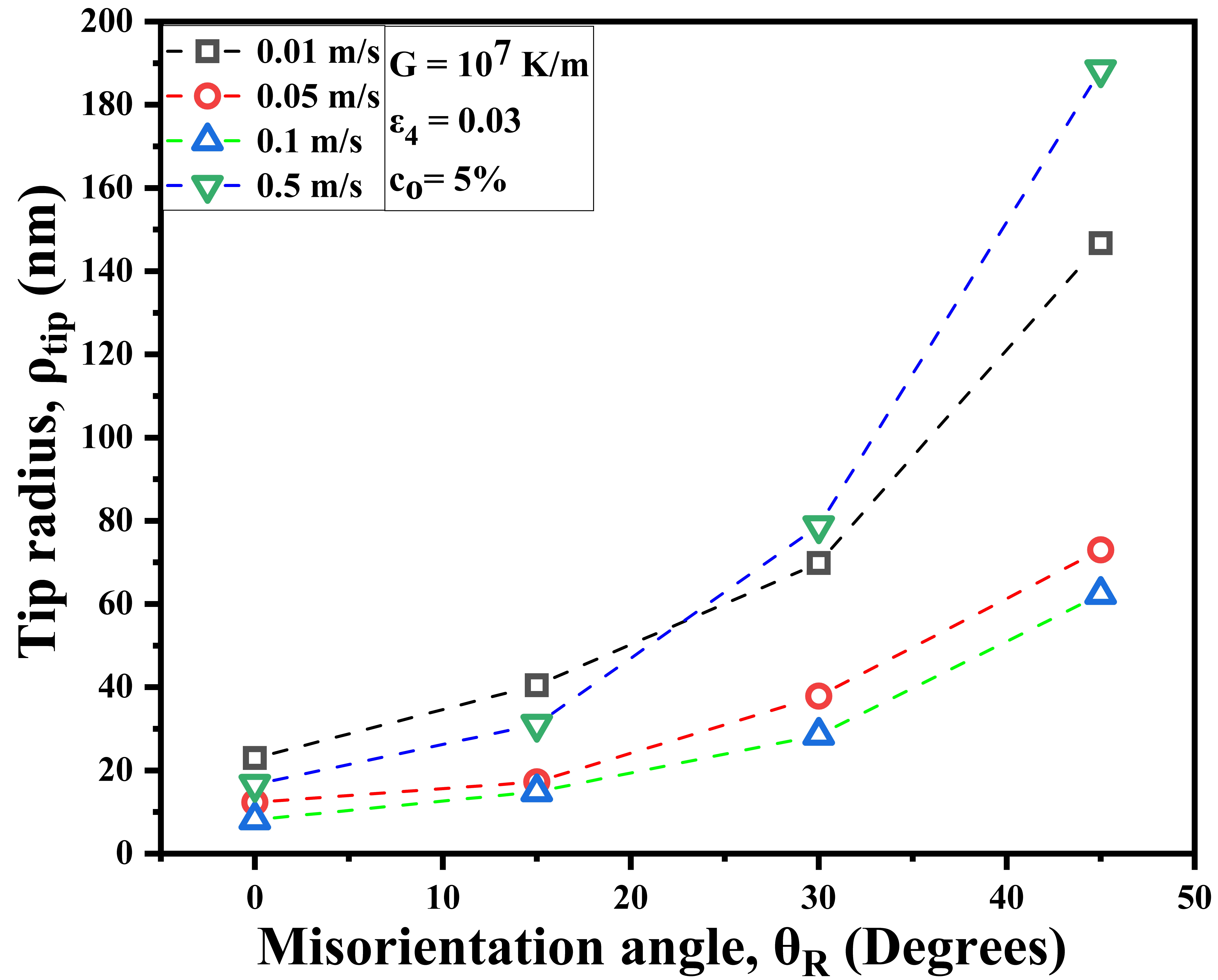}}\hfill
\subfloat[]{\includegraphics[width=0.33\textwidth]{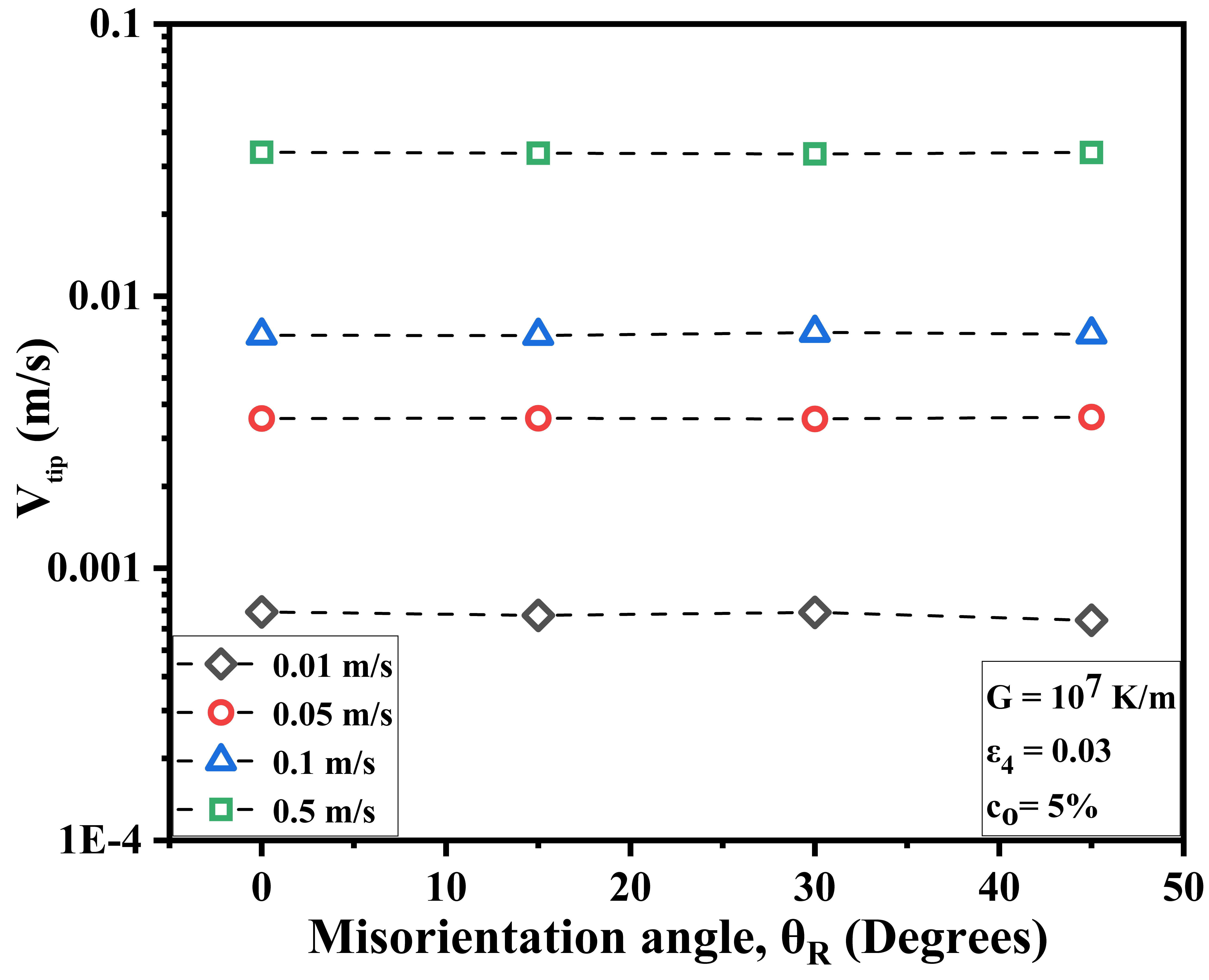}}\hfill
\subfloat[]{\includegraphics[width=0.33\textwidth]{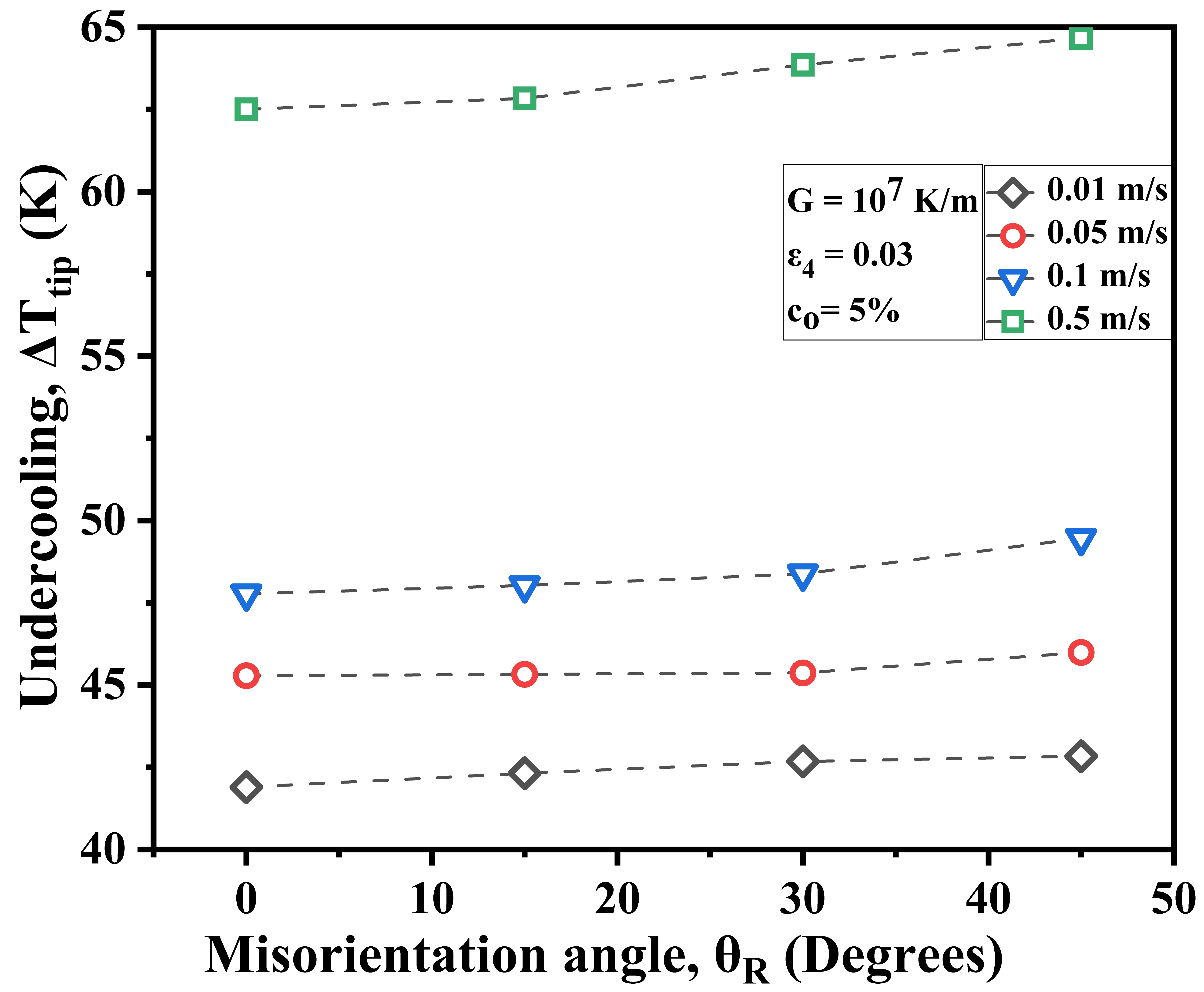}}
\caption{The variations in tip characteristics, (a) tip radius ($\rho_{tip}$), (b) tip velocity ($V_{tip}$), and (c) tip undercooling (${\Delta T}_{tip}$), are shown in the space of ($\theta_R$, $V$) for fixed $G = 10^7$ K/m. The fixed parameters and variable $V$ are shown in the inset.}\label{fig_tip}
\end{figure}

We obtain a spectrum of dendritic morphologies and transitions between them by varying $\theta_R$ in Fig.~\ref{fig_morphology}. The tip characteristics, such as the tip radius, growth speed, and undercooling, can distinguish these structures. An analysis for a parabolic tip of radius $\rho_{tip}$ as a function of $\theta_R$ is shown in Fig.~\ref{fig_tip}a. With increasing $\theta_R$, $\rho_{tip}$ increases as the interface progressively flattens due to the dendrite to cell transition as the condition for absolute velocity is approached or when the expected dendrite-to-seaweed transition occurs as a result of the growth competition. Similar observations have been reported at high velocity in~\cite{Boettinger1999}. We find that $\rho_{tip}$ increases by $\approx$5 to 10 times as $\theta_R$ increases from 0\degree to 45\degree. The tip region continues to become flatter over time with increasing $V$ (or decreasing $G$), eventually becoming a planar solidification front (\textit{i.e.}, $\rho_\text{tip}$ $\to$ $\infty$) beyond $V_{ab}$. This high-velocity planar regime will not be addressed here.

Furthermore, the $\theta_R$ has little influence on the leading tip velocity in the high-$V$ regime (see Fig.~\ref{fig_tip}b). This stands in contrast to previous studies on low-velocity solidification~\cite{xing2016}, when dendrites grew faster with a small $\theta_R$, resulting in less time for solute concentration to accumulate at the interface. These disparities in findings are not unexpected, given that our simulations are conducted close to the high-velocity cellular-planar instability limit, while the prior studies involved much lower velocities near the planar-cellular instability limit.

Undercooling is the thermodynamic driving force for solidification, and dendrites often grow faster in the direction of the largest undercooling. As solidification proceeds, the build-up of solute at the interface due to segregation increases the constitutional undercooling of the liquid. We estimate the tip undercooling ($\Delta T_{tip}$) by calculating the difference between the equilibrium liquidus temperature and the temperature at the leading dendrite tip. We plot the variation of $\Delta T_{tip}$ with $\theta_R$ for different values of $V$ in Fig.~\ref{fig_tip}c. Clearly, $\Delta T_{tip}$ increases with increasing $\theta_R$. When the variations of $G$ and $V$ are considered, $\Delta T_{tip}$ increases with increasing $V$ (or decreasing $G$) as well, consistent with previous studies~\cite{xing2016,xing2015phase}. 

An orientation competition exists between the seaweed and tilted dendrite in the transient states, predominantly at large $\theta_R$ (Fig.~\ref{fig_morphology}). The seaweed pattern results for 45\degree at $V$ = 0.5 m/s with a larger $\Delta T_{tip}$ compared to tilted dendrite, which forms at a lower $\theta_R$ with smaller $\Delta T_{tip}$. Thus, during the growth competition, seaweed cannot prevent the emergence of tilted growth, which dominates in the final microstructure (Fig.~\ref{fig_morphology}a, $V$ = 0.1 m/s). In contrast, the seaweed morphology is selected from the beginning stages of solidification for $V$ = 0.5 m/s and, hence, dominates in the final microstructure. Such growth orientation selection with respect to $\Delta T_{tip}$ suggests that dendrites are dynamically preferred at a lower $V$ with smaller $\Delta T_{tip}$ compared to the seaweed, which forms at a higher $V$ and $\theta_R$ with a larger $\Delta T_{tip}$. 

\subsection{Effects of anisotropy coefficient} 
\begin{figure}[ht]
\centering
\subfloat[$G = 10^7$ K/m, $V = 0.1$ m/s]{\includegraphics[width=0.4\textwidth]{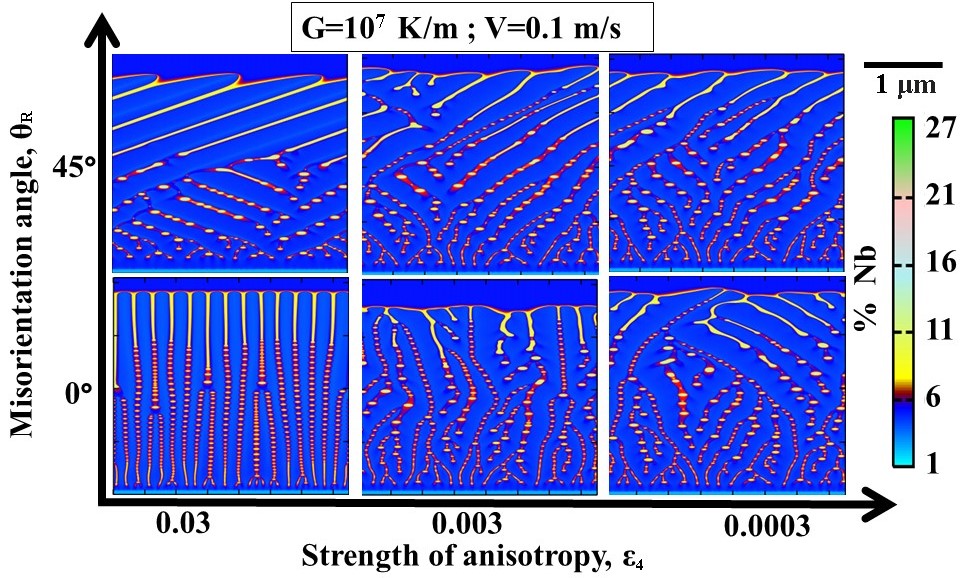}}\hfill
\subfloat[$G = 10^7$ K/m, $V = 0.5$ m/s]{\includegraphics[width=0.3\textwidth]{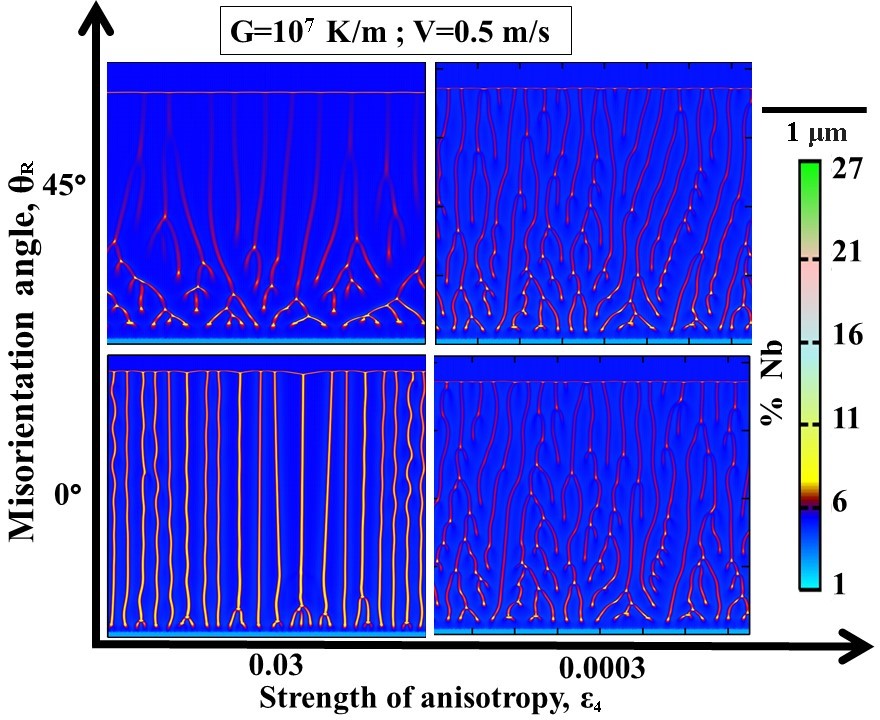}}\hfill
\subfloat[$G = 10^5$ K/m, $V = 0.5$ m/s]{\includegraphics[width=0.3\textwidth]{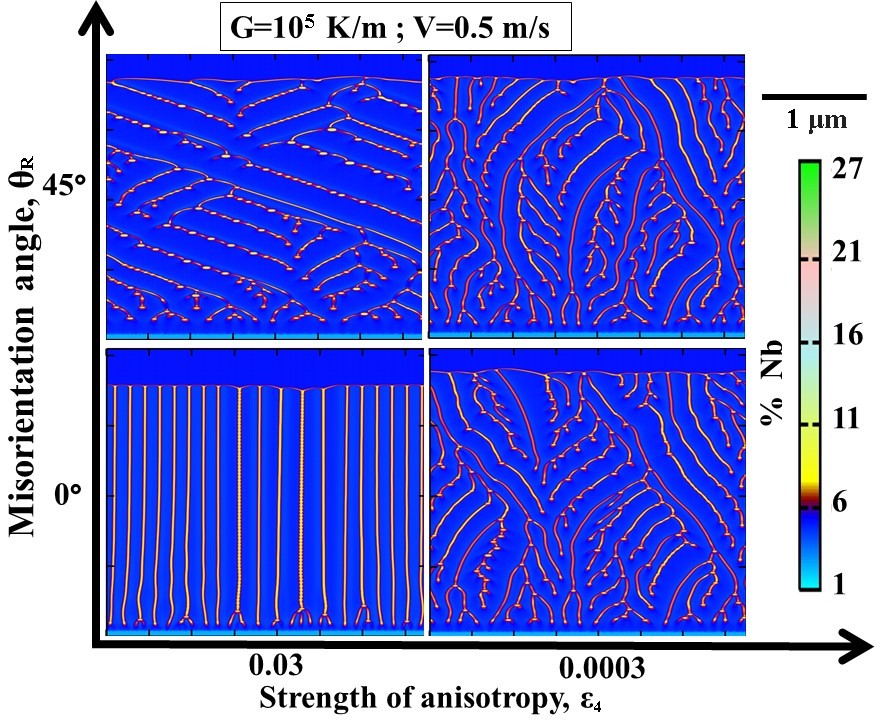}}
\caption{Morphology diagram for dendritic evolution in the space of ($\epsilon_4$, $\theta_R$). Selected combinations of ($G$, $V$, $\theta_R$) are used to best represent the simulation results obtained. The colorbar represents the solute profile. (For interpretation of the references to color in this figure legend, the reader is referred to the web version of this article.)}\label{fig_anisotropy}
\end{figure}

\begin{figure}[ht]
\centering
\includegraphics[scale=0.4]{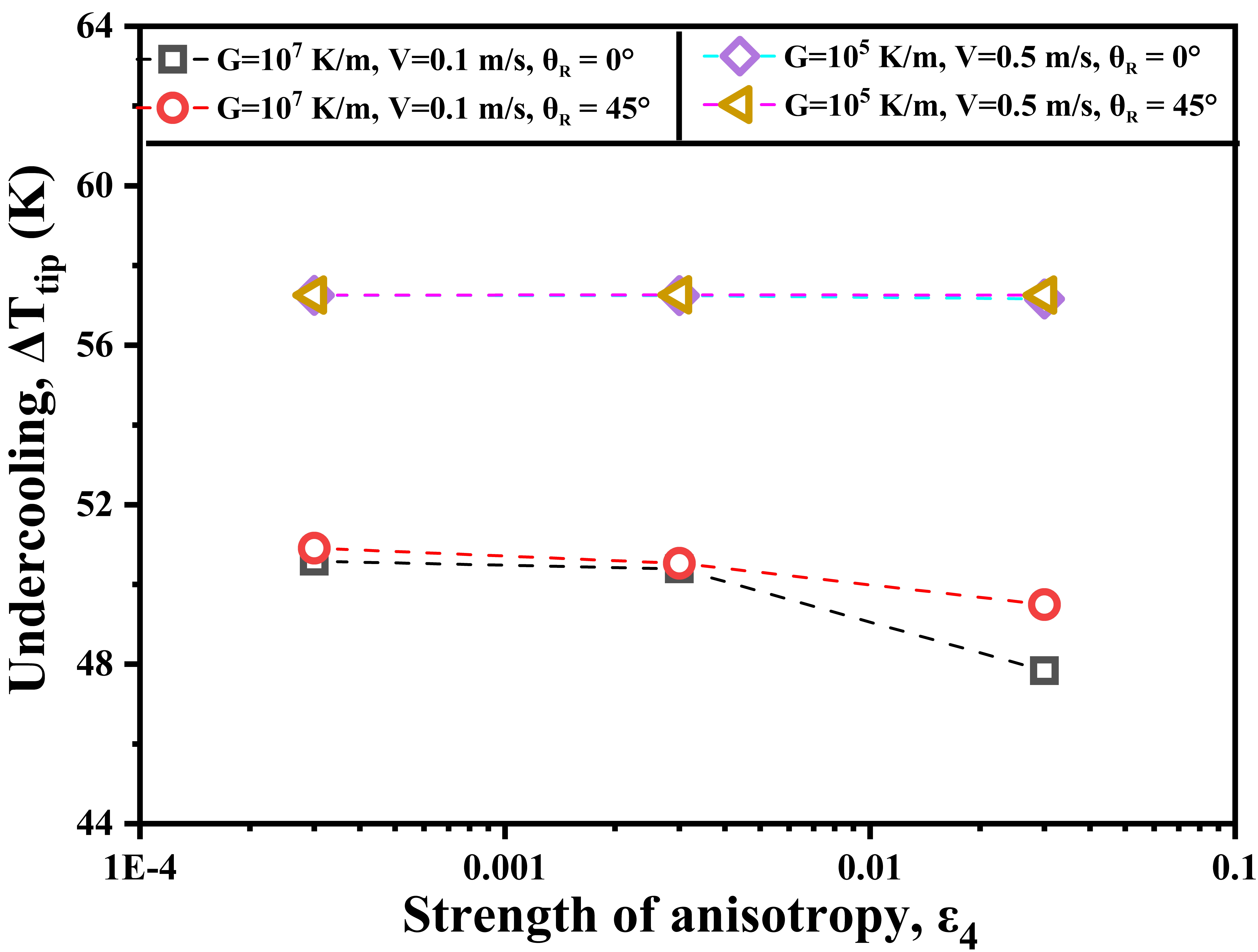}
\caption{Variation of the anisotropy parameter ($\epsilon_4$ in Eq.~\eqref{eq_anisotropy}) affects the leading tip undercooling. The plotted data for various ($G$, $V$, $\theta_4$) corresponds to the morphologies shown in Fig.~\ref{fig_anisotropy}.}\label{fig_aniso_analysis}
\end{figure}

Until now, we work with the orientation term of the effective anisotropy, that is, nonzero $\theta_R$ in Eq.~\eqref{eq_anisotropy}. For further investigation, we study the effects of the degree of effective anisotropy [$\epsilon_4$ in Eq.~\eqref{eq_anisotropy}] $\in$ [0.03, 0.0003]. We present results for the extreme values of $\theta_R$ (0\degree and 45\degree) and $V$ (0.1 m/s and 0.5 m/s) at fixed $G$ = $10^7$ K/m. The corresponding morphology diagram is shown in Fig.~\ref{fig_anisotropy}a. For $\theta_R$ = 0\degree at $V$ = 0.1 m/s, symmetrical dendrites parallel to axial direction develop for a strong anisotropy ($\epsilon_4$ = 0.03). For a small anisotropy ($\epsilon_4$ = 0.003), the interface morphology assumes a seaweed pattern consisting of doublons with sidebranches. Alternation of wide interdendritic grooves, delimiting the columnar seaweeds; narrow liquid grooves corresponding to doublons; and droplet ``pinch-off'' events from the groove bottoms are evident in this disordered morphology. These patterns are often generated by the alternating tip splitting instability close to the growth front, consistent with Fig.~14 in Ref.~\cite{akamatsu1998anisotropy}. When the anisotropy is further reduced ($\epsilon_4$ = 0.0003), the seaweed becomes strongly disordered, leading to a ``degenerate'' pattern, essentially made of competing dendritic and seaweed patterns tilted at arbitrary orientations to the left and the right of the $G$ direction~\cite{utter2001alternating,utter2002dynamics}. The characteristic features of degenerate dynamics are somewhat eliminated for a large $\theta_R$ (Fig.~\ref{fig_anisotropy}a, $\theta_R$ = 45\degree), resulting in a tilted or columnar seaweed pattern during low-anisotropy solidification. Since the degenerate morphology has a higher $\Delta T_{tip}$ (Fig.~\ref{fig_aniso_analysis}), it is likely to be overgrown by the tilted seaweed (tilted cell-like structure with a flat tip and lower $\Delta T_{tip}$) following a growth competition between these dendritic states, consistent with experimental observations~\cite{akamatsu1995symmetry,akamatsu1998anisotropy}. 

For $V$ = 0.5 m/s, the high-anisotropy solidification produces a columnar seaweed (Fig.~\ref{fig_anisotropy}b). Numerous cell termination and tip splitting-induced degeneracy become increasingly prominent with decreasing $\epsilon_4$. We do not observe any transient morphological transition to other dendrite states in the long-time-scale dynamics of dendritic evolution. Moreover, no droplet ``pinch-off'' events from the groove bottoms can be identified, implying that the crystal anisotropy may affect the formation of segregation-induced secondary phases in the solid material. For a relatively large $\theta_R$, we find that effects of decreasing $G$ (= $10^5$ K/m in Fig.~\ref{fig_anisotropy}c) are qualitatively equivalent to reducing the effective anisotropy, often leading to strongly disordered seaweed patterns with continuous tip splitting. The $\Delta T_{tip}$ does not change appreciably due to the morphological similarity that arises during this low-anisotropy solidification (Fig.~\ref{fig_aniso_analysis}). Similar observations have been reported in experiments~\cite{akamatsu1995symmetry,utter2001alternating}. 

\subsection{Effects of alloy composition}
Previous studies report complex interactions between melt concentration ($c_0$) and effective anisotropy ($\epsilon_4$ and $\theta_R$), including that the addition of solute amplifies the interfacial anisotropy at low $V$~\cite{haxhimali2006,dantzig2013dendritic}. Here, we simulate the effects of $c_0$ with respect to $\theta_R$ and $\epsilon_4$ for fixed thermal conditions ($G$ and $V$). We note that $V_{ab}$ of the alloy with 2.5 wt\%, 5 wt\%, and 7.5 wt\% Nb is given by 0.5 m/s, 1 m/s, and 2 m/s, respectively. During high-anisotropy solidification of the 2.5 wt\% alloy, the interface typically assumes an axially-oriented columnar seaweed morphology for $\theta_R$ = 0\degree and a planar front for $\theta_R$ = 45\degree. This planar front is consistent with $V_{ab}$ of the 2.5 wt\% alloy. Such morphological dissimilarity with respect to $\theta_R$ suggests that an orientation correction may be necessary for estimating $V_{ab}$ as $V_{ab}\, F(\Theta)$. More work is necessary to clarify this issue. The interface morphology becomes markedly different with increasing $c_0$. The interface develops a columnar seaweed for the 5 wt\% alloy and a tilted seaweed for the 7.5 wt\% alloy. Similar oriented structures result at a lower $V$ for the 5 wt\% alloy (Fig.~\ref{fig_morphology}a, $V$ = 0.1 m/s). This implies that the microstructure instabilities responsible for developing various dendrite patterns, including the dendrite-to-seaweed transition, shift to a higher $V$ (or smaller $G$) with increasing $c_0$. 

Next, we compare the interface morphologies that arise from low-anisotropy solidification ($\epsilon_4$ = 0.0003). A planar growth front results in the 2.5 wt\% alloy; a degenerate seaweed pattern develops in the 5 wt\% alloy; and a root-like dense branching seaweed consisting of many doublons arises in the 7.5 wt\% alloy. Such random appearance of dendritic evolution makes the characterization very challenging, hence not pursued here. Although not shown here, we obtain similar disordered microstructures in low-anisotropy alloys for $G$ = $10^5$ K/m (similar to Fig.~\ref{fig_anisotropy}c). The effects of $c_0$ are more pronounced in high-anisotropy alloys rather than low-anisotropy alloys, as evident in the analysis of tip undercooling in Fig.~\ref{fig_comp_analysis}. 

Until now, we computed orientation selection within a single dendrite grain characterized by a single global orientation. Finally, we perform a preliminary simulation to study competition between orientations (or crystals) of 0\degree and 30\degree in a bi-crystal setup. It should be noted that a change in the solidification pathway, including local solute microsegregation, modified the anisotropy parameter of the alloy material~\cite{haxhimali2006,akamatsu1995symmetry,wang2013large}. Therefore, the growth competition may not be just limited to the different orientation angles but can also be due to some range of anisotropy parameters between grains. Therefore, we also vary $\epsilon_4$ between grains in an ad-hoc manner. The qualitative resemblance between the morphologies in Fig.~\ref{fig_bicrystal} and Fig.~\ref{fig_experiment2}b is evident in that the tilted domains and disordered seaweed coexist after directional solidification in Ni-Nb alloys. In this approach, growth conditions for each crystal can be independently controlled, dramatically enlarging the scope of possible dendritic patterns and growth competition between them; a study of which is beyond the scope of this work and subject to another publication. Nevertheless, these observations imply the strong interactions between the crystal anisotropy and thermal conditions that may be at play during melt-pool solidification, and that these simulations can actually be seen as ``numerical experiments''.

\begin{figure}[ht]
\centering
\subfloat[$\epsilon_4 = 0.03$]{\includegraphics[width=0.5\textwidth]{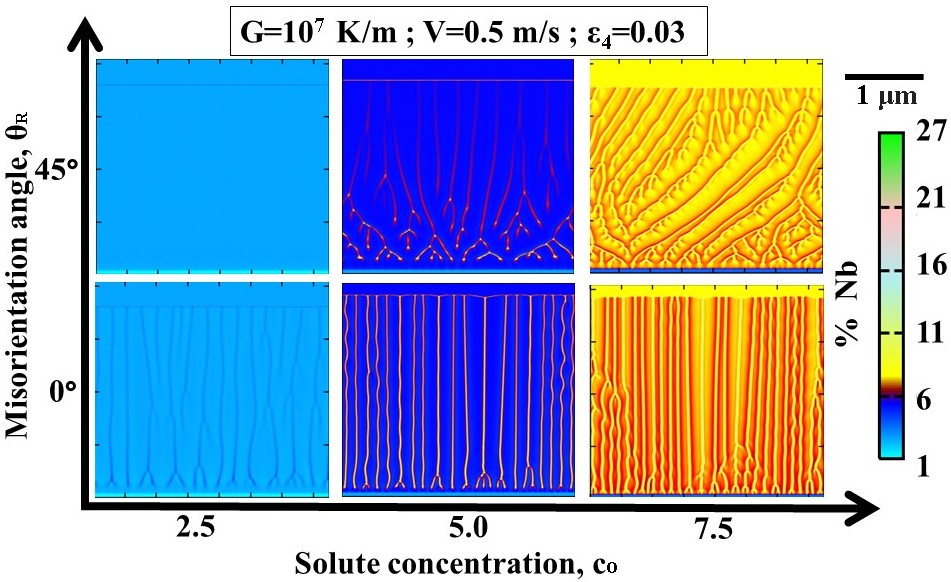}}\hfill
\subfloat[$\epsilon_4 = 0.0003$]{\includegraphics[width=0.5\textwidth]{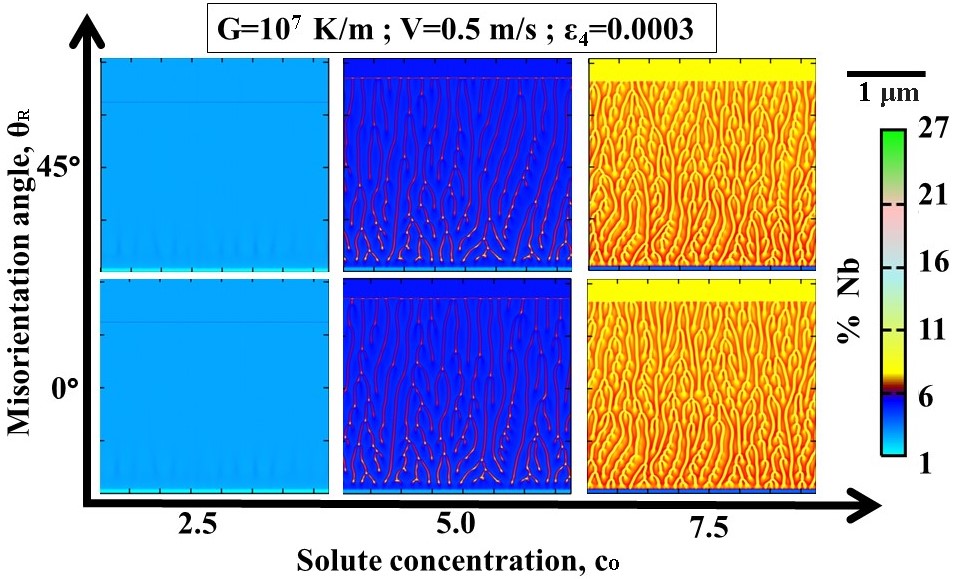}}
\caption{Morphological diagram for dendritic evolution in the space of ($c_0$, $\theta_R$). (a) High-anisotropy alloys with $\epsilon_4$ = 0.03 and (b) low-anisotropy alloys with $\epsilon_4$ = 0.0003. We present data for $G = 10^7$ K/m and $V$ = 0.5 m/s. The colorbar represents the solute (Nb) profile. (For interpretation of the references to color in this figure legend, the reader is referred to the web version of this article.)}\label{fig_composition}
\end{figure}

\begin{figure}[ht]
\centering
\includegraphics[scale=0.4]{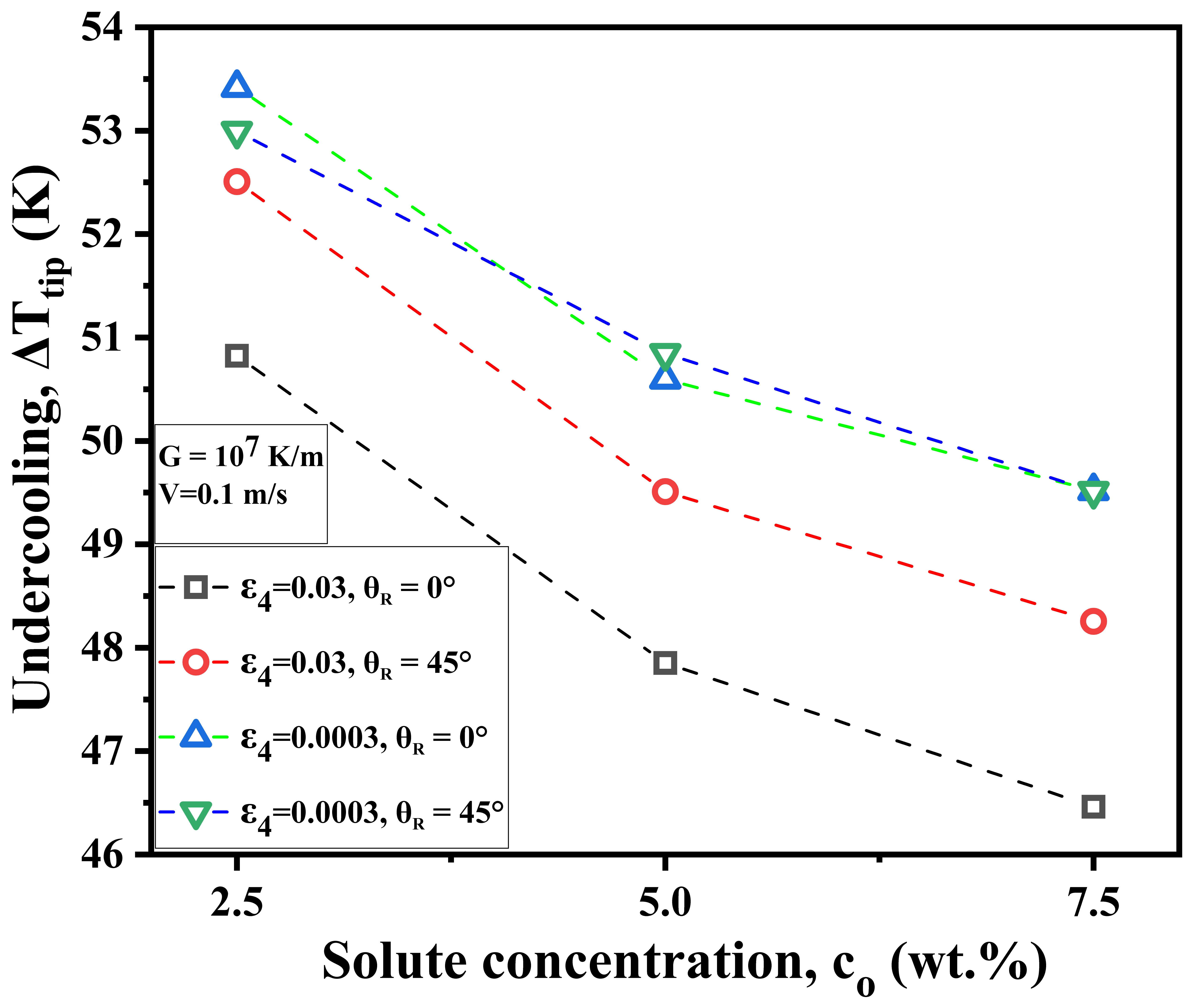}
\caption{Variation of the alloy composition ($c_0$) affects the leading tip undercooling. The plotted data in the space ($c_0$, $\epsilon_4$, $\theta_4$) corresponds to the morphologies shown in Fig.~\ref{fig_composition}.}\label{fig_comp_analysis}
\end{figure}

\begin{figure}[ht]
\centering
\includegraphics[width=0.6\textwidth]{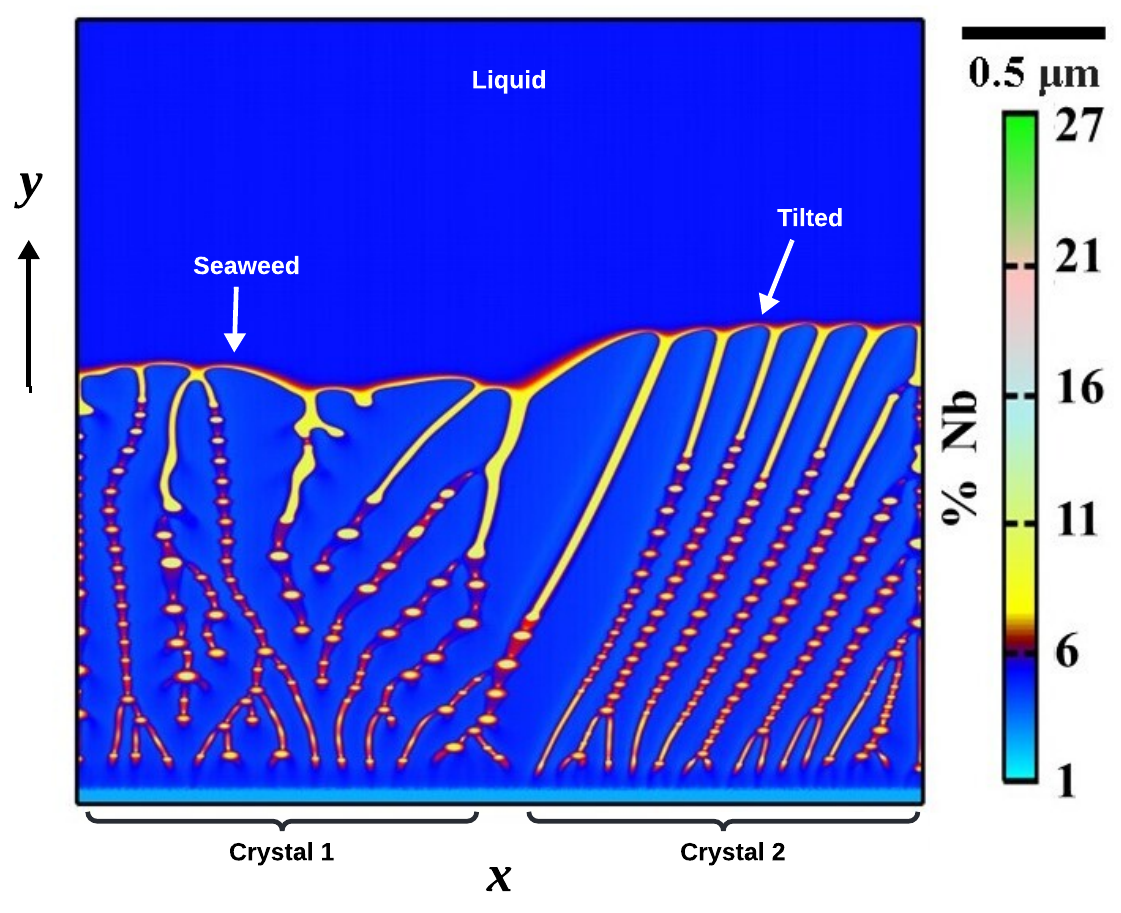}
\caption{Preliminary simulation of growth orientation competition in a cubic bi-crystal. The effective anisotropy in these crystals are different: ($\theta_R$ = 0\degree, $\epsilon_4$ = 0.0003) on the left, Crystal 1 and ($\theta_R$ = 30\degree, $\epsilon_4$ = 0.03) on the right, Crystal 2. The low- and high anisotropy parameters on the left and right crystals select a morphology consisting of seaweed (left) and tilted (right) structures, similar to the melt-pool microstructure shown in Fig.~\ref{fig_experiment2}b. We use a no-flux boundary condition in all directions. The colorbar represents the Nb profile. (For interpretation of the references to color in this figure legend, the reader is referred to the web version of this article.)}\label{fig_bicrystal}
\end{figure}

\section{Discussion}\label{sec_discussion}
To answer the question of orientation selection at high velocity, we have used two-dimensional phase-field simulations of dendritic evolution in a broad parameter space ($\theta_R$, $V$, $G$, $\epsilon_4$, and $c_0$) near the high velocity (absolute stability) condition. We consider the example of AM processing, where the inclination angles of the melt-pool boundary generate different solidification sites with misorientation angle $\theta_R$, leading to different dendrite growth directions~\cite{tao2019_kv}. We use a phase-field model that has been benchmarked by detailed numerical tests in two-dimensions and has produced quantitative agreement with reference experiments at low $V$. Here, we work with high $V$ representative of AM, where complex physical interactions involving melt convection and interface kinetics may be present, which we ignore as a first approximation. The extremely fine scales (average of $\lambda < \SI{1.5}{\micro \meter}$) and the dilute alloy consideration (5 wt\% Nb) minimize the convection effects on the compositions~\cite{ghosh2017primary,ghosh2023_review}. Therefore, our simulations give at least a preliminary account of dendrite orientations at high $V$. Of course, for a more quantitative analysis, the interface kinetics and curvature corrections, among others, must be considered to model rapid solidification, noting the various growth directions and curvilinear dendrites that may arise under different $\theta_R$. Despite these limitations, some useful observations arise out of our simulations. 

Our simulations demonstrate that dendrite morphology within a grain typically changes in the following sequence: symmetrical dendrite $\rightarrow$ tilted dendrite $\rightarrow$ columnar seaweed $\rightarrow$ degenerate seaweed, as $\theta_R$ increases from 0\degree to 45\degree for given growth conditions. The crystalline anisotropy begins to affect the dendrite orientation as $V$ approaches $V_{ab}$. We find that for a relatively large $\theta_R$, increasing $V$ is qualitatively equivalent to decreasing the anisotropy, which can explain the occurrence of the tilted dendrite-to-seaweed transition and vice versa. Small anisotropies lead to different kinds of seaweed morphologies, including columnar, tilted, and degenerate, sometimes combined with doublons, as observed in experiments~\cite{utter2001alternating,utter2002dynamics}. To our knowledge, the selection mechanisms of seaweed patterns remain unknown in AM. We demonstrate here that despite their limited stability, they have important implications for the solidification pathways, microsegregation patterns, and the segregation-induced secondary phases, and hence could influence the occurrence of solidification defects, particularly the hot tearing cracks in the solidified material~\cite{debroy_additive,sun2022hottearing}. In some instances, a discrete morphology rather than the continuous chain morphology of the secondary phase has been found to minimize the crack density by altering the local residual stress states~\cite{nie_2014}. The solute-rich droplets that are detached from the interdendritic grooves (similar to our simulations) could provide nucleation sites in experiments for the formation of secondary phases at the end of solidification; thus, the droplet trajectory variation observed in simulations may have implications for their morphology in the solid material. We find that with decreasing anisotropy, the distribution of droplets changes from a continuous network to an isolated pattern due to the growth competition between two dendritic states. A finite value of $\theta_R$ may be necessary to minimize the degenerate dynamics of the seaweed morphology, which could transform into anisotropy-directed tilted dendrites or axially-oriented columnar seaweed depending on the growth condition. We find that droplet formation could be avoided at high $V$ close to $V_{ab}$. This may have implications for developing microstructures free from harmful secondary phases in the final material.

While previous studies~\cite{haxhimali2006,dgp_2008,akamatsu1998anisotropy,xing2016} have investigated the growth dynamics of tilted dendrites at low $V$, to our knowledge, this is the first systematic study of orientation selection at high $V$, with various combinations of the control parameters. The qualitative picture that emerges from our simulations is consistent with experiments and previous numerical studies. The primary spacing, constitutional undercooling, and segregation ratio increase with increasing $\theta_R$ and $V$ ($\rightarrow$ $V_{ab}$). These dendrite features are compared with the commonly regarded theories~\cite{dgp_2008,gandin1996orientation} involving orientation correction for the tilted growth. However, these scaling theories do not apply well to our data. This is not surprising given the different alloy systems and thermal conditions employed in previous studies, or maybe it is because these theories are not appropriate at high $V$ but apply to well-developed tilted arrays at a low $V$. Moreover, most of these studies do not consider the tilted growth at 45\degree, the largest misorientation for a cubic material.

During the transient growth competition between different dendrite states, the pattern with lower $\Delta T_{tip}$ under the same growth condition often dominates in the final microstructure. Thus, we find that tilted dendrites are dynamically preferred over seaweed within a grain, similar to the previous studies~\cite{akamatsu1998anisotropy,xing2016}. Also, some of the interface morphologies that we obtain here are very similar to experimental observations at low $V$~\cite{akamatsu1995symmetry,akamatsu1998anisotropy}, although for different alloy and solidification parameters. These observations imply that the morphological and orientation selections are more of an anisotropy-driven phenomenon rather than that of $V$. Conversely, the dendrite length scale ($\lambda$), microsegregation ($k_V$), and the distribution of secondary phases primarily depend on $V$. We find that a transition from tilted dendrites to columnar/degenerate seaweed occurs with increasing $V$ and decreasing anisotropy, as reported previously for anisotropy-driven dynamics of cellular fronts at low $V$~\cite{akamatsu1998anisotropy}. Finally, we observe a planar growth front of a very different nature from columnar seaweed for $V \geq V_{ab}$ (Fig.~\ref{fig_composition}, 2.5 wt\% alloy), demonstrating the high-$V$ stabilization of the planar interface as predicted by Mullins and Sekerka and observed experimentally~\cite{Mullins1964}.

It is clear that anisotropy-directed tilted dendrite/texture predominantly results in AM. Such preferred growth directions often lead to undesired effects of anisotropy in the mechanical properties of the solidified product~\cite{kok_2018,kouraytem2021}. This kind of anisotropy could be useful for some applications, such as single crystal turbine blades, requiring good high-temperature creep strength properties. In contrast, mitigating strong columnar texture can be beneficial for applications, for example, requiring good fatigue strength properties. Therefore, even if the undesired columnar structures can not be avoided entirely in AM, texture randomization \textit{via} the seaweed-type morphology has good potential for microstructure control. The seaweed morphologies contributed to better resistance to dislocation motion in experiments, leading to better plasticity and work hardening effects in the solidified material~\cite{wang2022enhanced,xing2020}. The above observations can aid in tailoring the optimum available microstructure for each section of the solidified product in terms of local performance requirements, \textit{e.g.}, fatigue and fracture resistance at one location and creep resistance at another~\cite{moat2009crystallographic}. We demonstrate that by solute ($c_0$) additions and adjusting the growth conditions ($G$ and $V$), which control the effective anisotropy ($\epsilon_4$ and $\theta_R$) of the crystalline material, the orientation selection mechanism in dendritic evolution can be designed and controlled.

\section{Conclusions and outlook}\label{sec_conclusions}
The selection mechanisms of tilted dendrite and seaweed morphologies need to be better understood in the frame of AM solidification. Therefore, we use a phase-field model~\cite{Echebarria2004} to simulate the dendrite morphology and growth orientation selection at high velocity. We have used a relatively large control parameter space to study the effects of varying $G$, $V$, $\theta_R$, $\epsilon_4$, and $c_0$. From these simulations, we highlight the following observations:

\begin{itemize}
\item The $\theta_R$ significantly affects the interface morphology. For $\theta_R$ = 0\degree, symmetrical dendrites grow along the thermal gradient direction. For a nonzero $\theta_R$, tilted growth of dendrite arrays results. For a sufficiently large $\theta_R$ (45\degree), a seaweed dendritic structure often results with no specific growth direction angle.

\item For a fixed $\theta_R$, we find that tilted dendrites form at a low $V$. With increasing $V$, a growth competition between the tilted and seaweed dendrites can be seen. A seaweed morphology develops for a sufficiently large $V$. Overall, we find that the occurrence of dendrite-to-seaweed transition is promoted by increasing $V$ (or decreasing $G$), increasing $\theta_R$, and decreasing $\epsilon_4$. We demonstrate that $\Delta T_{tip}$ offers insight into the effect that competing preferred growth directions have on the orientation that dendrites eventually select.

\item A large $\theta_R$ or $V$ can effectively eliminate the characteristic features of degenerate dynamics in the seaweed morphology. This can explain the emergence of the patches of preferred orientations that may arise within an otherwise disordered morphology in the solidified melt-pool.

\item We find that the characteristic dendritic features, including the primary arm spacing, microsegregation, and liquid constitutional undercooling, increase with increasing $\theta_R$ under the same growth condition.

\item In high-anisotropy solidification, tilted dendrites often result at a low $\theta_R$ while seaweeds dominate at a relatively large $\theta_R$. The tip profiles (curvature and undercooling) effectively change with increasing $\theta_R$. In low-anisotropy solidification, various seaweed structures develop, ranging from columnar to tilted to degenerate. The columnar seaweed nearly follows the thermal gradient direction, while the degenerate lacks the apparent preferred orientation of the crystal.

\item We find that the commonly accepted theories regarding dendrite growth direction at low $V$ do not apply well to dendrite orientations obtained at high $V$. As an AM-Benchmark study~\cite{lyle2020outcomes}, we have presented the values of the fitting parameters in these empirical relationships, which link growth direction angle to relevant variables at high $V$.

\item The solute partitioning across the interfacial region significantly varies with $\theta_R$ and $V$, with important implications for microsegregation, solute trapping, and the distribution of interdendritic secondary phases in the solid material.

\item The trajectory of the solute-rich droplets that are shed from the interdendritic liquid form continuous chain patterns into the solid at small $\theta_R$, while large $\theta_R$ leads to isolated, discrete patterns of the same. Such growth dynamics may influence the morphology and distribution of the secondary phases, with important implications for the hot cracking sensitivity of the dendrite material.
\end{itemize}

The above observations will guide more detailed numerical simulations, particularly in three-dimensions~\cite{ghosh20183d,ghosh_isotherm}, and address some key open questions. Which is relatively more dominating parameter for controlling the interface morphology at high $V$: the $V$ itself or the effective anisotropy? How does the addition of solute elements (\textit{e.g.}, bcc Fe or hcp Ti in fcc Ni-Nb alloys regarding Inconel 718) affect the orientation selection at high $V$? Strongly tilted seaweeds reveal a twofold rather than fourfold anisotropy in experiments~\cite{utter2002dynamics}; this situation needs verification to check for morphological similarity in the simulated systems~\cite{ghosh2017_eutectic}. Moreover, multigrain simulations in which the interplay of various $\theta_R$ during the convergent or divergent growth conditions in bi-crystals will contribute to more physical insights on oriented dendrites in real materials~\cite{tourret_2015,xing2020}. Work in this particular direction is currently in progress. The interdendritic segregation-induced secondary phases often assume a eutectic morphology in the solid~\cite{kang2022_cellular_eutectic,rafieazad2019effects}; thus, the orientation competition in a hybrid eutectic-dendritic morphology at high velocity in low-anisotropy alloys would be very interesting~\cite{ghosh2015_pre,ghosh2017_eutectic}. Finally, the possibility of altering $\theta_R$ by adjusting the process parameters in AM will provide new prospects to control the final morphology, length scale, orientation, and non-equilibrium phases in a dendrite material, and even enable the design of site-specific dendrite directionality (\textit{i.e.}, texture control) to suit a given application~\cite{moat2009crystallographic}.

%----------------------------------------------------------------------------------------

%% If you have acknowledgments, this puts in the proper section head.
\section*{Acknowledgments}
The authors acknowledge the National Supercomputing Mission for providing computing resources of PARAM Ganga at IIT Roorkee implemented by the Center for Development of Advanced Computing (C-DAC) and supported by the Ministry of Electronics and Information Technology and Department of Science and Technology, Government of India. S. Ghosh acknowledges the support of the Faculty Initiation Grant from the Sponsored Research \& Industrial Consultancy Office, Indian Institute of Technology Roorkee, and Science and Engineering Research Board, Government of India.

\section*{Conflict of Interest Statement}
The authors have no conflicts to disclose.

\section*{Data Availability Statement}
The data that supports the findings of the study are available from the corresponding author upon reasonable request.

\section*{References}
%\bibliography{papers}

\end{document}